\documentclass[twocolumn,superscriptaddress,floatfix]{revtex4-2}

\RequirePackage{graphicx}

\usepackage{amsmath}
\usepackage{amssymb}
\usepackage{booktabs}
\usepackage{float}
\usepackage{gensymb}
\usepackage{graphicx}
\usepackage{hyperref}
\usepackage{placeins}
\usepackage{soul}
\usepackage{upgreek}
\usepackage{xcolor}
\usepackage{subcaption}
\usepackage{xfrac}
\usepackage{mhchem}
\usepackage{lineno}

\begin{document}

\title{Demonstration of neutron identification in neutrino interactions in the MicroBooNE liquid argon time projection chamber}
% List of institutions in command form:
\newcommand{\ANL}{Argonne National Laboratory (ANL), Lemont, IL, 60439, USA}
\newcommand{\Bern}{Universit{\"a}t Bern, Bern CH-3012, Switzerland}
\newcommand{\BNL}{Brookhaven National Laboratory (BNL), Upton, NY, 11973, USA}
\newcommand{\UCSB}{University of California, Santa Barbara, CA, 93106, USA}
\newcommand{\Cambridge}{University of Cambridge, Cambridge CB3 0HE, United Kingdom}
\newcommand{\CIEMAT}{Centro de Investigaciones Energ\'{e}ticas, Medioambientales y Tecnol\'{o}gicas (CIEMAT), Madrid E-28040, Spain}
\newcommand{\Chicago}{University of Chicago, Chicago, IL, 60637, USA}
\newcommand{\Cincinnati}{University of Cincinnati, Cincinnati, OH, 45221, USA}
\newcommand{\CSU}{Colorado State University, Fort Collins, CO, 80523, USA}
\newcommand{\Columbia}{Columbia University, New York, NY, 10027, USA}
\newcommand{\Edinburgh}{University of Edinburgh, Edinburgh EH9 3FD, United Kingdom}
\newcommand{\FNAL}{Fermi National Accelerator Laboratory (FNAL), Batavia, IL 60510, USA}
\newcommand{\Granada}{Universidad de Granada, Granada E-18071, Spain}
\newcommand{\Harvard}{Harvard University, Cambridge, MA 02138, USA}
\newcommand{\IIT}{Illinois Institute of Technology (IIT), Chicago, IL 60616, USA}
\newcommand{\Imperial}{Imperial College London, London SW7 2AZ, United Kingdom}
\newcommand{\Indiana}{Indiana University, Bloomington, IN 47405, USA}
\newcommand{\KSU}{Kansas State University (KSU), Manhattan, KS, 66506, USA}
\newcommand{\Lancaster}{Lancaster University, Lancaster LA1 4YW, United Kingdom}
\newcommand{\LANL}{Los Alamos National Laboratory (LANL), Los Alamos, NM, 87545, USA}
\newcommand{\Louisiana}{Louisiana State University, Baton Rouge, LA, 70803, USA}
\newcommand{\Manchester}{The University of Manchester, Manchester M13 9PL, United Kingdom}
\newcommand{\MIT}{Massachusetts Institute of Technology (MIT), Cambridge, MA, 02139, USA}
\newcommand{\Michigan}{University of Michigan, Ann Arbor, MI, 48109, USA}
\newcommand{\MSU}{Michigan State University, East Lansing, MI 48824, USA}
\newcommand{\Minnesota}{University of Minnesota, Minneapolis, MN, 55455, USA}
\newcommand{\Nankai}{Nankai University, Nankai District, Tianjin 300071, China}
\newcommand{\NMSU}{New Mexico State University (NMSU), Las Cruces, NM, 88003, USA}
\newcommand{\Oxford}{University of Oxford, Oxford OX1 3RH, United Kingdom}
\newcommand{\Pitt}{University of Pittsburgh, Pittsburgh, PA, 15260, USA}
\newcommand{\QMUL}{Queen Mary University of London, London E1 4NS, United Kingdom}
\newcommand{\Rutgers}{Rutgers University, Piscataway, NJ, 08854, USA}
\newcommand{\SLAC}{SLAC National Accelerator Laboratory, Menlo Park, CA, 94025, USA}
\newcommand{\SDSMT}{South Dakota School of Mines and Technology (SDSMT), Rapid City, SD, 57701, USA}
\newcommand{\Maine}{University of Southern Maine, Portland, ME, 04104, USA}
\newcommand{\Syracuse}{Syracuse University, Syracuse, NY, 13244, USA}
\newcommand{\TelAviv}{Tel Aviv University, Tel Aviv, Israel, 69978}
\newcommand{\Tennessee}{University of Tennessee, Knoxville, TN, 37996, USA}
\newcommand{\UTA}{University of Texas, Arlington, TX, 76019, USA}
\newcommand{\Tufts}{Tufts University, Medford, MA, 02155, USA}
\newcommand{\UCL}{University College London, London WC1E 6BT, United Kingdom}
\newcommand{\VTech}{Center for Neutrino Physics, Virginia Tech, Blacksburg, VA, 24061, USA}
\newcommand{\Warwick}{University of Warwick, Coventry CV4 7AL, United Kingdom}
%\newcommand{\Yale}{Wright Laboratory, Department of Physics, Yale University, New Haven, CT, 06520, USA}
%%\newcommand{\listerThanks}{Now at University of Wisconsin, Madison}

% So that institutions appear in alphabetical order:
\affiliation{\ANL}
\affiliation{\Bern}
\affiliation{\BNL}
\affiliation{\UCSB}
\affiliation{\Cambridge}
\affiliation{\CIEMAT}
\affiliation{\Chicago}
\affiliation{\Cincinnati}
\affiliation{\CSU}
\affiliation{\Columbia}
\affiliation{\Edinburgh}
\affiliation{\FNAL}
\affiliation{\Granada}
\affiliation{\Harvard}
\affiliation{\IIT}
\affiliation{\Imperial}
\affiliation{\Indiana}
\affiliation{\KSU}
\affiliation{\Lancaster}
\affiliation{\LANL}
\affiliation{\Louisiana}
\affiliation{\Manchester}
\affiliation{\MIT}
\affiliation{\Michigan}
\affiliation{\MSU}
\affiliation{\Minnesota}
\affiliation{\Nankai}
\affiliation{\NMSU}
\affiliation{\Oxford}
\affiliation{\Pitt}
\affiliation{\QMUL}
\affiliation{\Rutgers}
\affiliation{\SLAC}
\affiliation{\SDSMT}
\affiliation{\Maine}
\affiliation{\Syracuse}
\affiliation{\TelAviv}
\affiliation{\Tennessee}
\affiliation{\UTA}
\affiliation{\Tufts}
\affiliation{\UCL}
\affiliation{\VTech}
\affiliation{\Warwick}
%\affiliation{\Yale}

% Authors in alphabetical order
\author{P.~Abratenko} \affiliation{\Tufts}
\author{O.~Alterkait} \affiliation{\Tufts}
\author{D.~Andrade~Aldana} \affiliation{\IIT}
\author{L.~Arellano} \affiliation{\Manchester}
\author{J.~Asaadi} \affiliation{\UTA}
\author{A.~Ashkenazi}\affiliation{\TelAviv}
\author{S.~Balasubramanian}\affiliation{\FNAL}
\author{B.~Baller} \affiliation{\FNAL}
\author{A.~Barnard} \affiliation{\Oxford}
\author{G.~Barr} \affiliation{\Oxford}
% ...

\collaboration{The MicroBooNE Collaboration}
\thanks{microboone\_info@fnal.gov}\noaffiliation
%\email[]{microboone\_info@fnal.gov}

\date{\today}

\begin{abstract}
A significant challenge in measurements of neutrino oscillations is reconstructing the incoming neutrino energies.
While modern fully-active tracking calorimeters such as liquid argon time projection chambers in principle allow the measurement of all final state particles above some detection threshold, undetected neutrons remain a considerable source of missing energy with little to no data constraining their production rates and kinematics.
We present the first demonstration of tagging neutrino-induced neutrons in liquid argon time projection chambers using secondary protons emitted from neutron-argon interactions in the MicroBooNE detector.
We describe the method developed to identify neutrino-induced neutrons and demonstrate its performance using neutrons produced in muon-neutrino charged current interactions.
The method is validated using a small subset of MicroBooNE's total dataset.
The selection yields a sample with $60\%$ of selected tracks corresponding to neutron-induced secondary protons.
\end{abstract}

\maketitle

\section{Introduction}

With the study of neutrino oscillations moving into a precision measurement era with the Deep Underground Neutrino Experiment (DUNE) and Hyper-Kamiokande (Hyper-K) experiments under construction \cite{DUNE_TDR, HK_TDR}, significant work is still needed to reduce the uncertainties that arise due to the modeling of neutrino-nucleus interactions.
As neutrino energies are typically estimated based on the visible final state particles produced in their interactions (e.g. \cite{MINOS_osc, NOvA_osc, uBooNE_osc}), the presence of neutrons will bias these estimates even for fully active detectors such as the liquid argon time projection chamber (LArTPC) technology that DUNE will use.
While developments have been made in understanding and improving neutrino energy resolutions using new methods \cite{FurmanskiSobczyk, FurmanskiHilgenberg, DL_EE}, neutrons will always represent a bias that requires model-dependent corrections.
To first order, the neutrino's reconstructed energy will be biased by the full kinetic energy of the neutron.
While neutrons produced in neutrino interactions can be observed through capture processes, particularly in gadolinium-loaded detectors \cite{SuperK, ANNIE}, this technique is not easily applicable to LArTPC detectors and provides little information other than the presence of a neutron.
Neutrino-induced neutrons can also be measured through inelastic scatters with nuclei, for example using liquid scintillator detectors (as demonstrated by the COHERENT experiment \cite{COHERENT}) or in solid scintillator detectors (e.g. in the MINERvA experiment \cite{Minerva2019, Minerva2023}).
In the case of the MINERvA experiment, these measurements were made using antineutrino interactions which tend to produce more neutrons.
It has also been shown using the ArgoNeuT detector that neutrons can be identified in a LArTPC through the photons produced by the de-excitations of excited argon nuclei produced in neutron inelastic scatters \cite{Acciarri_2019}.
This work introduces a method of identifying neutrino-induced neutrons in a liquid argon TPC using neutron-argon inelastic scatters that produce a secondary proton, similar to the technique demonstrated by MINERvA.
The method is validated using data from the MicroBooNE detector.

\subsection{MicroBooNE Detector}
The MicroBooNE LArTPC detector is a $2.56$\,($x$)\,$\times$\,$2.33$\,($y$)\,$\times$\,$10.37$\,($z$)\,m$^{3}$, 85 tonne active volume LArTPC exposed to the Booster Neutrino Beam (BNB) at Fermi National Accelerator Laboratory that took data between 2015-2021, with an exposure of $1.49 \times 10^{21}$ protons on target (POT) \cite{uBooNE_JINST}.
This work uses a subset of that data, corresponding to an exposure of $4.13 \times 10^{19}$\,POT delivered in 2017-2018.
The MicroBooNE detector is located $468$ m downstream of the target, at which point the neutrino beam primarily consists of $\nu_{\mu}$ ($93.7\%$) with a mean energy of $\mathtt{\sim}$800\,MeV \cite{MB_flux}.
Charged particles produced in neutrino-argon interactions propagate through the LArTPC and ionize the argon atoms.
The ionized charge drifts in a 273\,V/cm electric field towards the anode that consists of two induction planes and one collection plane, with 3\,mm pitch between wires and 3 mm plane separation \cite{SCE_microboone}.
The full drift time for an electron in MicroBooNE is 2.3\,ms.
There is an average of six cosmogenic particles in the detector during a drift window \cite{CosmicPaper, PMT_Paper}.
The LArTPC is paired with 32 photo-multiplier tubes (PMTs) that identify scintillation photons from argon atoms excited by particles traversing the liquid argon.
By leveraging the time and spatial structure of the scintillation light, and comparing to the visible TPC activity, cosmogenic backgrounds that are out of time with the 1.6\,$\mu$s BNB spill can be efficiently tagged and removed.

\subsection{MicroBooNE Simulation}
The MicroBooNE collaboration uses software that employs a series of Monte Carlo simulations that model all aspects of the experiment.
The BNB flux simulation was originally developed by the MiniBooNE collaboration \cite{MB_flux} and then customized for MicroBooNE.
The neutrino-argon interactions are modeled in the MicroBooNE simulation chain using the GENIE v3.0.6 neutrino event generator \cite{ANDREOPOULOS201087, AlvarezRuso} with a modified version of the G18\textunderscore10a\textunderscore02\textunderscore11a tune.
This modified tune of G18\textunderscore10a\textunderscore02\textunderscore11a was developed to better fit the 2016 CC0$\pi$ T2K dataset \cite{PhysRevD.105.072001}.
Final state particles from the interactions simulated by GENIE are propagated through the MicroBooNE detector using GEANTv4\textunderscore10\textunderscore3\textunderscore p03c \cite{GEANT4, GEANT4_recent_developments}.
The physics package used in the MicroBooNE version of GEANT4 is the quark-gluon string model and the Bertini cascade model (QGSP\textunderscore BERT) \cite{GEANT4}.
Data taken while the BNB is off is overlaid on the simulated neutrino interaction to describe both the cosmic background and intrinsic detector noise.
The detector response is modeled in the LArSoft framework \cite{Snider_2017}.

The neutrons of interest in this work are produced in neutrino interactions.
An additional source of neutrons comes from secondary interactions of neutrino-induced hadrons within the detector volume.
These neutrons are distinct from the primary neutrons in that they originate from a different nucleus.
However, their signatures are indistinguishable and they constitute an irreducible background.

\begin{figure*}[htb!]
\centering
\includegraphics[width=.5\linewidth]{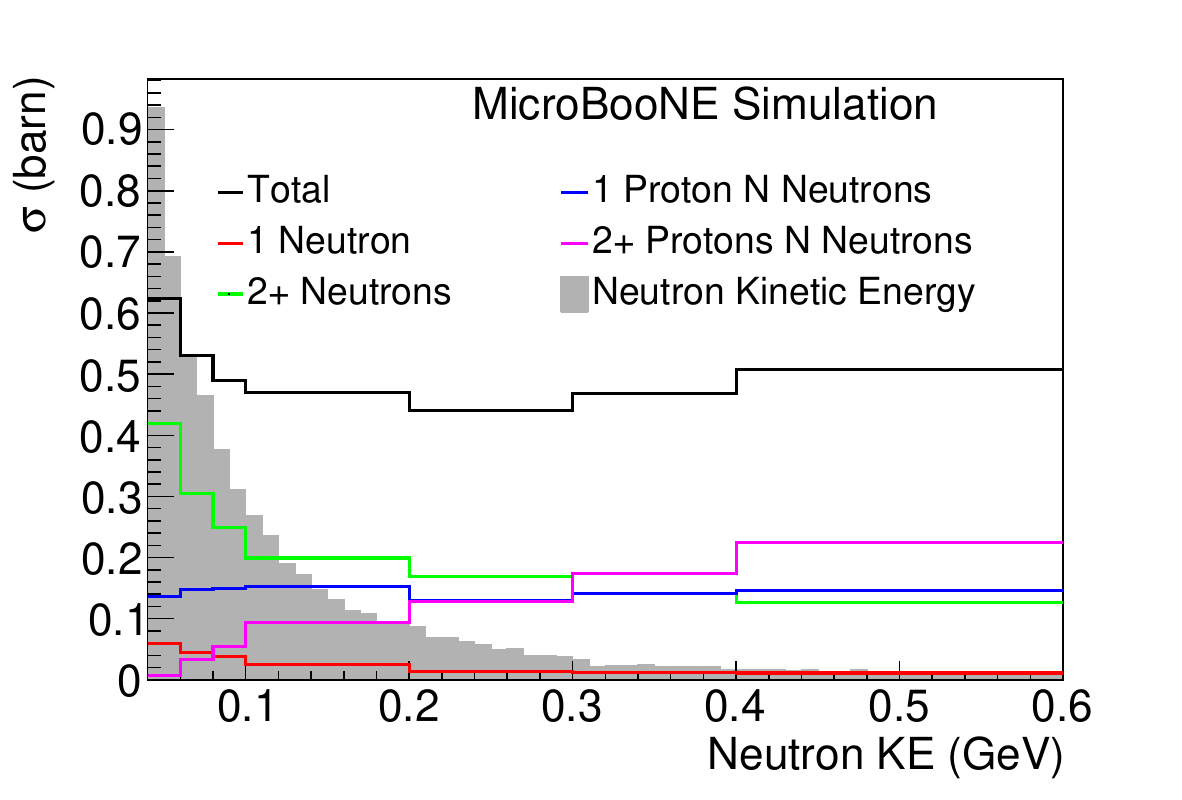}
\caption{Neutron-argon inelastic scattering cross section as a function of neutron kinetic energy and broken down by the most common inelastic scatter final states. The predicted neutron energy spectrum is shown in gray, with arbitrary normalization.
}
\label{fig:neutron_argon_xsec}
\end{figure*}

\section{Neutron Detection}
In order to tag and identify neutrons in neutrino interactions on argon, it is important to understand their behavior as they propagate through the detector.
This section outlines the characteristics of neutron inelastic scattering on argon.
%The neutron-argon total cross section between neutron kinetic energies of 95 and 720\,MeV is fairly flat and approximately 700\,mb \cite{MiniCaptain2023}.
%This cross section corresponds to a mean free path in liquid argon of $\mathtt{\sim}$70\,cm.

Tables~\ref{table:NeutronEndProcess}-~\ref{table:NeutronFinalState} summarize the behavior of neutrons in the MicroBooNE detector, according to the GEANT4 simulation deployed.
Importantly, of the neutrons in our sample, $71.45\%$ inelastically scatter at least once within the detector before leaving the active volume.
The majority of neutrons inelastically scatter only once ($30.28\%$) or twice ($26.51\%$).
More than half ($54.13\%$) of neutron inelastic scatters result in only the original neutron and the possibly excited secondary argon atom, while another $30.86\%$ of scatters produce additional neutrons.
Finally, $8.85\%$ of neutron-argon inelastic scatters create protons, with $6.60\%$ of scatters including one proton and $1.37\%$ of scatters including two protons.

For BNB neutrino-induced neutrons in MicroBooNE, the flux-integrated cross section is dominated by the 1 neutron and 2 or more neutron final states, due to the low average neutron energies.
Figure ~\ref{fig:neutron_argon_xsec} shows the neutron-argon inelastic scattering cross section (as extracted from the GEANT4 simulation) as a function of the neutron kinetic energy and broken down by the final states listed in Table~\ref{table:NeutronFinalState}.
The cross section is only shown for neutron energies above 40~MeV, the rough minimum to which our analysis method – reconstruction of a secondary proton track – is sensitive.
%the region where this method is sensitive to neutrons---neutron kinetic energies between 40 and 600\,MeV.
The energy spectrum predicted by the GENIE simulation for neutrons produced by neutrinos in MicroBooNE is shown in gray, with arbitrary normalization.
These theoretical predictions have not been directly validated with data \cite{LEPLAR}, although in the energy range of interest the total (including elastic interactions) neutron-argon cross section has been measured \cite{MiniCaptain2023}.
The neutron-argon total cross section between neutron kinetic energies of 95 and 720\,MeV is fairly flat and approximately 700\,mb, corresponding to a mean free path in liquid argon of $\mathtt{\sim}$70\,cm.

In the energy range of interest (between 40 and 600\,MeV), the proton production cross section is approximately flat.
The detection method presented here targets secondary protons produced by inelastic scatters, represented by the pink and blue lines in Fig~\ref{fig:neutron_argon_xsec}.  Proton-generating processes become the dominant predicted exclusive interaction channel above roughly 300 MeV in neutron kinetic energy.
A similar method has been shown to work to identify neutrons in the ProtoDUNE detector \cite{RiveraThesis}.

\begin{table}[ht]
\centering
\caption{Final state neutron end processes.}
\begin{tabular}{ |p{5.5cm}|p{1.1cm}|}
 \hline
 Neutron End Process&\\
 \hline
 Exit detector without inelastic scatter&$24.63\%$\\
 Exit detector after inelastic scatter&$71.45\%$\\
 Captured by argon&$3.13\%$\\
 Other&$0.75\%$\\
 \hline
\end{tabular}
\label{table:NeutronEndProcess}
\end{table}

\begin{table}[ht]
\centering
\caption{Number of times final state neutrons inelastically scatter.}
\begin{tabular}{ |p{4.cm}|p{1.1cm}|}
 \hline
 Number of Inelastic Scatters&\\
 \hline
 0 scatters&$25.48\%$\\
 1 scatter&$30.28\%$\\
 2 scatters&$26.51\%$\\
 3 scatters&$11.87\%$\\
 4+ scatters&$5.87\%$\\
 \hline
\end{tabular}
\label{table:NeutronScatter}
\end{table}

\begin{table}[ht]
\centering
\caption{Final state particles of neutron-argon inelastic scatters.}
\begin{tabular}{ |p{4.cm}|p{1.1cm}|}
 \hline
 Final State of Scatter&\\
 \hline
 (1) $\ce{Ar}$*&$54.13\%$\\
 (2) $\ce{Ar}$* $+ Nn$&$30.86\%$\\
 (3) $\ce{Cl} + Nn + 1p$&$6.60\%$\\
 (4) $\ce{S}+Nn+2p$&$1.37\%$\\
 (5) $\ce{S}+Nn+\alpha$&$2.14\%$\\
 (6) Other proton final state&$0.88\%$\\
 (7) Other&$4.03\%$\\
 \hline
\end{tabular}
\label{table:NeutronFinalState}
\end{table}

\begin{figure*}[htb!]
\centering
\includegraphics[width=.6\linewidth]{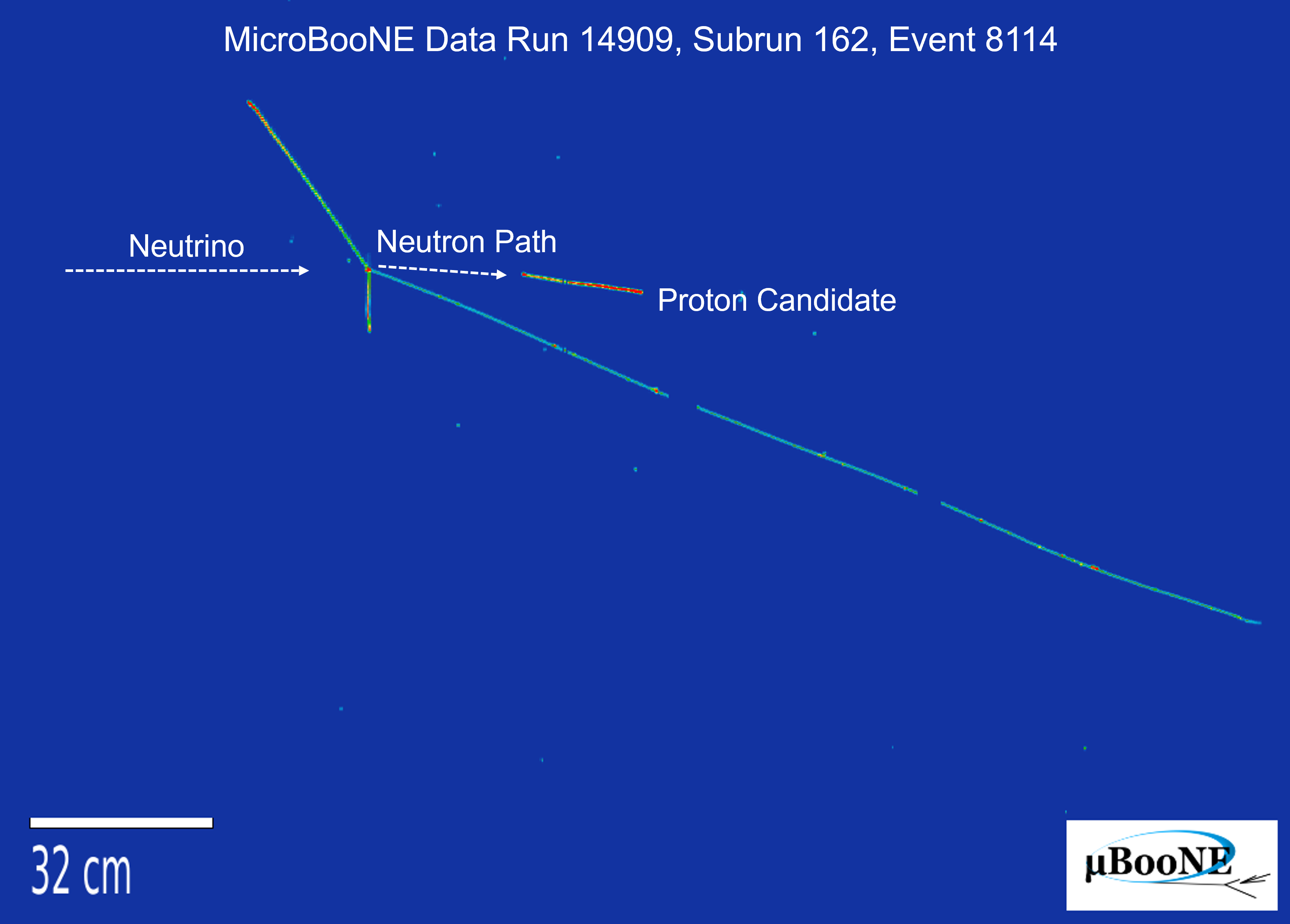}
\caption{Data event showing $\nu_{\mu}$ CC candidate with a final state neutron candidate. A secondary proton track candidate is seen in the middle of the image. Image is of the collection plane. Here the horizontal axis shows the z direction and the vertical axis shows the x direction in the MicroBooNE detector. Charge deposited is represented by the color scale, where blue corresponds to small charge deposits and red corresponds to large charge deposits. }
\label{fig:Event Display}
\end{figure*}

\begin{figure*}[htb!]
    \centering
        \includegraphics[width=.5\linewidth]{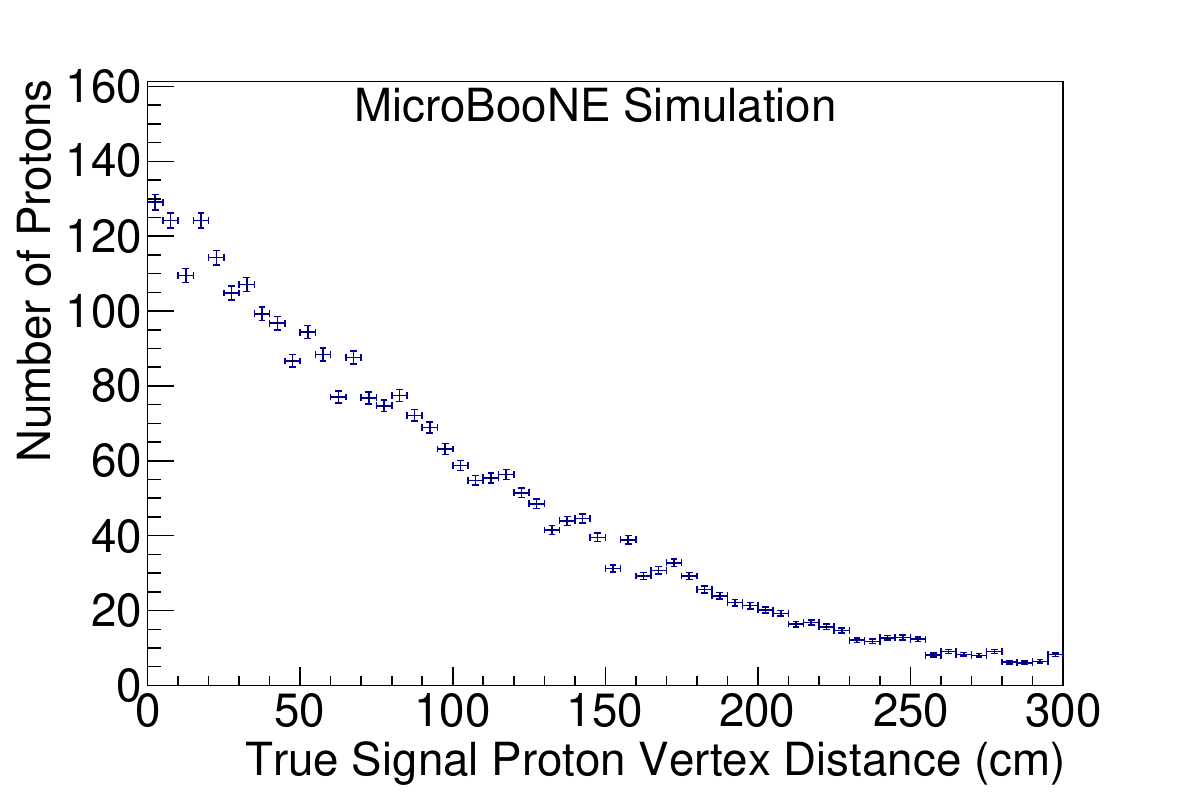}
        \caption{Neutrino vertex displacement for true secondary protons created from final state neutrons. Error bars shown here only represent statistical error.}
        \label{fig:true_p_vtx_disp}
\end{figure*}

\begin{figure*}[t!]
    \centering
    \begin{subfigure}[t]{0.49\textwidth}
        \centering
        \includegraphics[width=1.\linewidth]{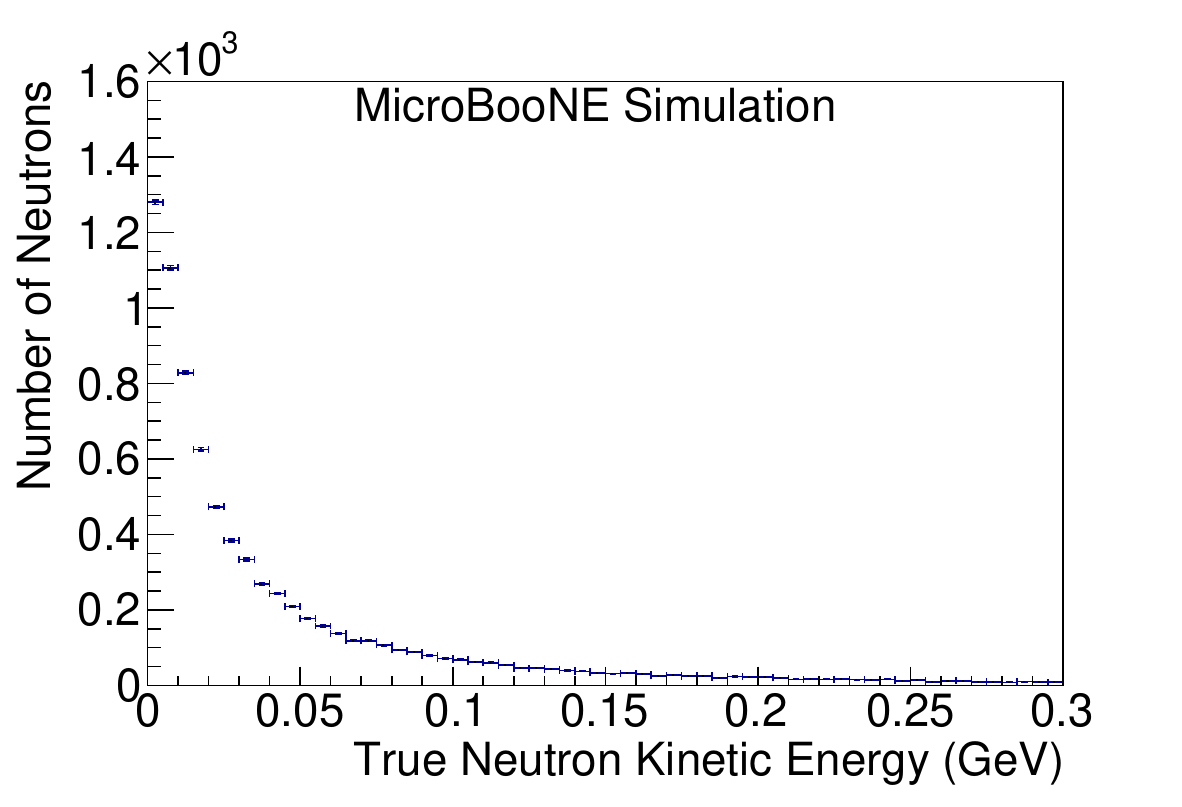}
        \caption{}
        \label{fig:true_n_KE_reduced_x_range}
    \end{subfigure}%
    ~ 
    \begin{subfigure}[t]{0.49\textwidth}
        \centering
        \includegraphics[width=1.\linewidth]{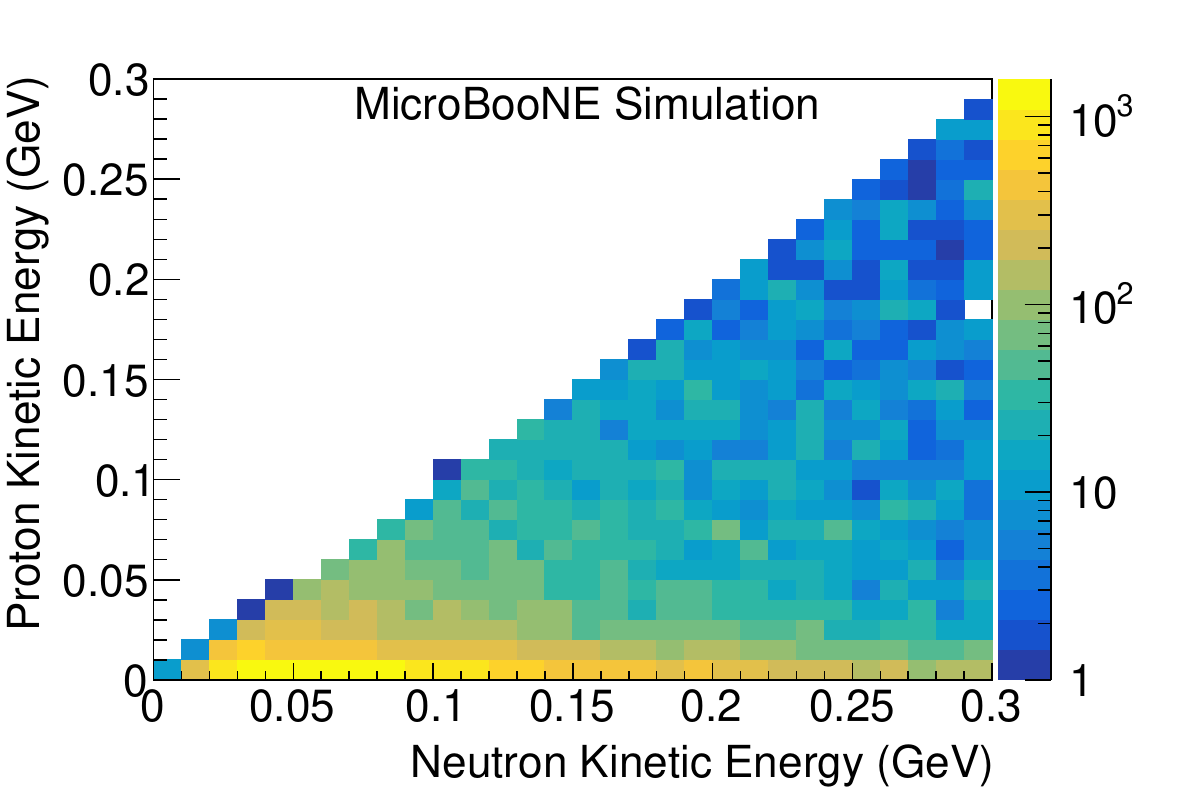}
        \caption{}
        \label{fig:n_sp_energy_transfer}
    \end{subfigure}
    \caption{(a) Kinetic energy spectrum for all $\nu_{\mu}$ CC final state true neutrons. (b) Correlation between parent neutron kinetic energy and daughter proton kinetic energy made from MicroBooNE simulation. Error bars shown here only represent statistical error.}
\end{figure*}

\begin{figure*}[t!]
    \centering
    \begin{subfigure}[t]{0.49\textwidth}
        \centering
        \includegraphics[width=1.\linewidth]{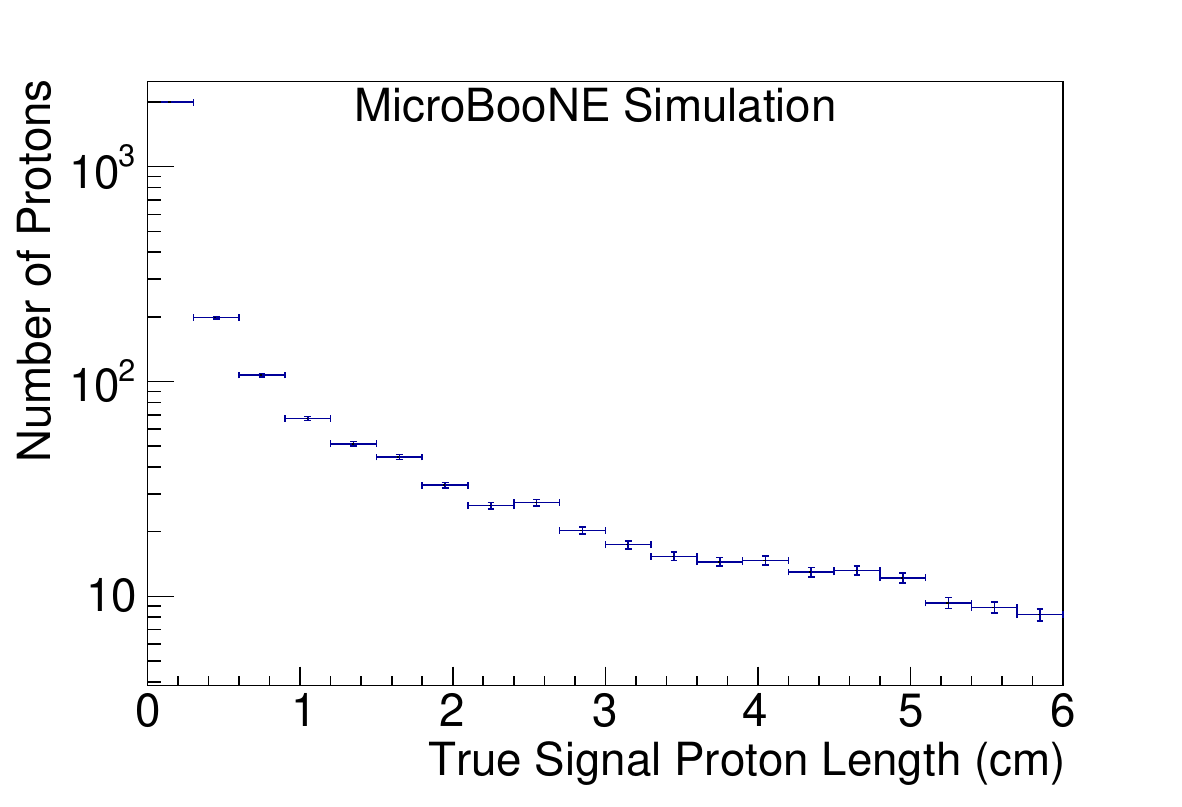}
        \caption{}
        \label{fig:true_p_len}
    \end{subfigure}%
    ~ 
    \begin{subfigure}[t]{0.49\textwidth}
        \centering
        \includegraphics[width=1.\linewidth]{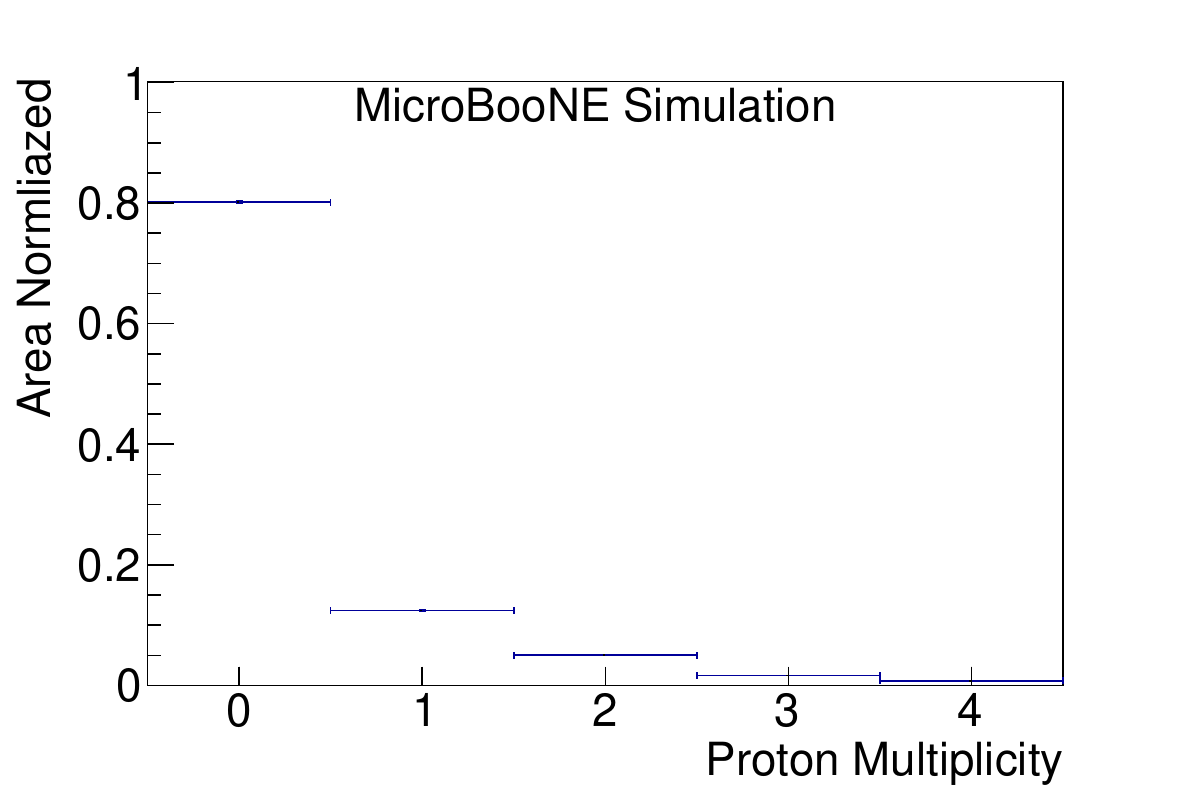}
        \caption{}
        \label{fig:protons_produced}
    \end{subfigure}
    \caption{(a) Track length of true secondary protons created from final state neutrons.  (b) Area normalized number of protons produced by final state neutrons. Error bars shown here only represent statistical error.}
\end{figure*}

The most frequent type of neutron interaction leaves the argon nucleus in an excited state without producing another visible signature.
In principle, MicroBooNE has the LArTPC detector response attributes and existing low-energy reconstruction tools \cite{RadonPaper, BiPoPaper} to tag neutrons through de-excitation photons.
While previous demonstrations of identifying such events have been produced \cite{Acciarri_2019}, these interactions present unique challenges that complicate their application to the problem of neutrino calorimetry \cite{FriedlandLiEnergyRes}.
%these interactions make it difficult to use such interactions to recover meaningful information to improve the calorimetric reconstruction of the neutrino energy.
When leveraging low energy photons, there is a contaminating population of photons from de-excitation of the remnant nucleus from the initial neutrino interaction, as well as a large amount of cosmogenic and radiogenic low energy EM activity, especially in a surface LArTPC such as the MicroBooNE detector.
In addition, the isotropic nature of these de-excitation photons, along with a 14\,cm radiation length in LAr, means any directional information which could help constrain the neutron's kinematics is significantly smeared \cite{Acciarri_2019, FriedlandLiContainment}.
The detection method presented here targets instead the secondary protons produced by inelastic scatters.
While this interaction mode is less common, these events present fewer problematic backgrounds, and the secondary proton location can be used to better estimate the neutron direction.
Additionally, secondary protons will be most sensitive to higher energy neutrons which are the most relevant when considering potential biases to neutrino energy estimation.

It is worth noting that the number of neutrons that produce visible protons depends on the detector geometry.
With the MicroBooNE detector being less than 2.5 m in dimensions of width and height, all neutrino interactions are less than 2 neutron interaction lengths from the nearest detector boundary.
A larger detector, such as those being constructed for DUNE, will therefore have a higher efficiency for tagging neutrons.

Figure~\ref{fig:Event Display} shows a MicroBooNE data event display that includes a charged current (CC) candidate interaction (upper left of image) with a final state neutron candidate. This neutron re-scatters on an argon nucleus and produces the proton track candidate (red track in the center of the image). Position information from this track allows for the reconstruction of the direction of the neutron, as well as the distance between the secondary proton and the neutrino interaction vertex.

Identifying neutrons via secondary protons presents certain challenges.
The exponential distribution in Fig.~\ref{fig:true_p_vtx_disp} shows that many secondary protons appear distant from the true neutrino vertex.
Most neutrons produced in neutrino interactions from the BNB are very low energy (Fig.~\ref{fig:true_n_KE_reduced_x_range}) and because neutrons transfer a small amount of kinetic energy to the daughter protons, there is minimal correlation between the neutron's kinetic energy and the secondary proton's kinetic energy (Fig.~\ref{fig:n_sp_energy_transfer}).
This results in short proton tracks which are hard to identify.
Figure~\ref{fig:true_p_len} shows the true secondary proton track length distribution from simulation. 
The majority of the protons ($\mathtt{\sim}$74\%) created from final state neutrons travel less than 3\,mm, meaning that they can cross at most 1 wire from a single plane in the detector. 
Figure~\ref{fig:protons_produced} shows the number of protons produced by a final state neutron, showing that 70\% of neutrons do not produce any secondary protons at all.
This apparent disagreement with the results shown in Table~\ref{table:NeutronFinalState} is because neutrons can scatter multiple times, thus providing more opportunities to produce a proton.
Though this identification method leads to a very low efficiency for low energy neutrons, it does identify higher-energy neutrons at a higher efficiency.

\section{Event Selection}

The reconstruction of charged particle tracks is performed by the Pandora multi-algorithm pattern recognition toolkit \cite{microboonecollaboration2017Pandora} which we briefly describe here.
Hits on the three planes are passed to the Pandora cosmic reconstruction to first identify unambiguous cosmic-ray muons and their daughter delta rays.
Hits associated with these backgrounds are removed from the input hit collection. This hit collection is subsequently passed to the Pandora neutrino reconstruction.
Some cosmic tracks---the ones that are not obviously of cosmic origin---are deliberately reconsidered in the next stage of the reconstruction to avoid possible loss of neutrino-induced tracks at an early stage in the process.
The remaining hits are divided into groups, called slices, that Pandora classifies as related using proximity and direction-based metrics with the goal of isolating the various interactions in the detector, whether induced by a neutrino or cosmic-ray.
Cosmic-ray-oriented and neutrino-oriented clustering and topological algorithms are separately applied to the hits in each slice, so that the two outcomes can be compared. The most appropriate outcome is identified for each slice.
Collections of hits within each slice are reconstructed to form particle flow particles (PFPs)---a general term that can refer to either a shower or a track---and a topological score is calculated for each slice using a support vector machine (SVM).
Slices resembling a neutrino interaction get a score closer to 1 and slices resembling cosmic activity get a score closer to 0.
Based on this score and confirming that the candidate neutrino slice has a flash that is in time with the BNB spill, at most one neutrino slice candidate per event is chosen.
The slicing process is especially relevant for the reconstruction of neutrons, since they regularly scatter far away from the neutrino vertex and daughter protons are often not included in the neutrino slice.

We begin our preselection with the sample of events identified as $\nu_{\mu}$ CC as in \cite{GardinerNote} where the neutrino vertex is contained in the fiducial volume:
\begin{center}
\begin{minipage}{2in}
\begin{itemize}
\item[] 21.50\,cm $\leq x \leq$ 234.85\,cm,
\item[] -95.00\,cm $\leq y \leq$ 95.00\,cm,
\item[] 21.50\,cm $\leq z \leq$ 966.80\,cm.
\end{itemize}
\end{minipage}
\end{center}
We additionally require all candidate tracks to be fully contained in a ``containment volume'' \cite{GardinerNote}:
\begin{center}
\begin{minipage}{2.5in}
\begin{itemize}
    \item[] 10.00\,cm $\leq$ x $\leq$ 246.35\,cm,
    \item[] -106.50\,cm $\leq$ y $\leq$ 106.50\,cm,
    \item[] 10.00\,cm $\leq$ z $\leq$ 1026.80\,cm.
\end{itemize}
\end{minipage}
\end{center}
We then consider all reconstructed PFPs within the neutrino slice to search for secondary protons from primary neutrons.  
For the purpose of studying the selection's performance, we group these into a variety of categories based on their origin:

\begin{figure*}[htb!]
\centering
\begin{subfigure}{.98\textwidth}
    \centering
    \includegraphics[width=.9\linewidth]{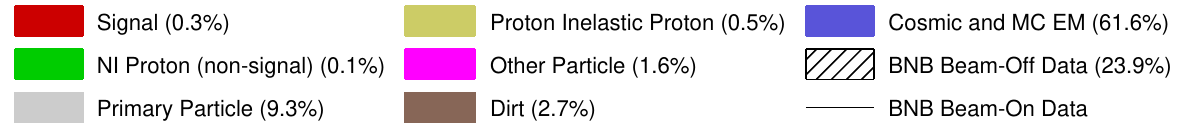}
\end{subfigure}
\begin{subfigure}{0.49\textwidth}
    \includegraphics[width=\linewidth]{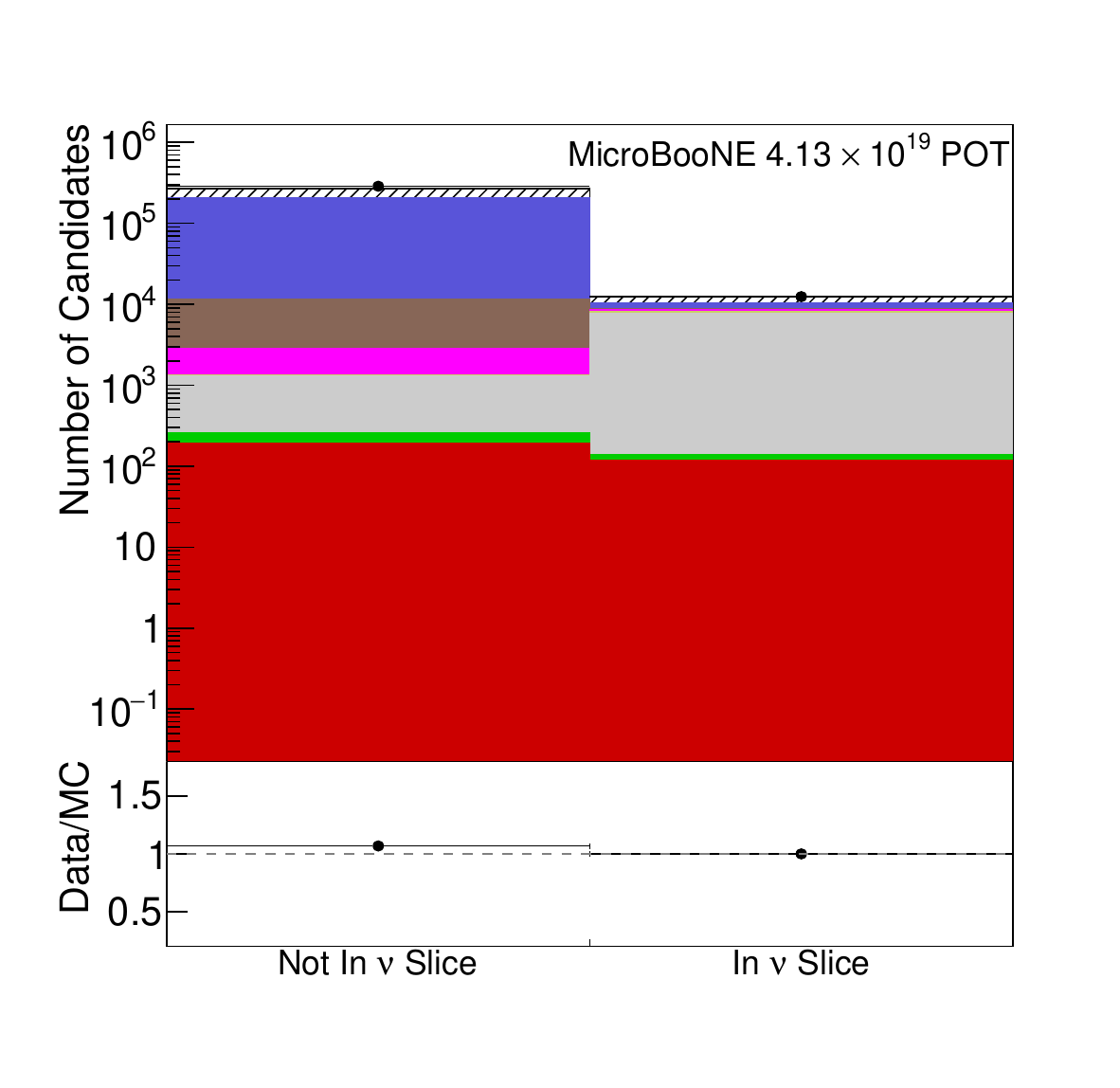}
    %\caption{}
    %\label{fig:run3_pfp_slc_presence_preselection}
\end{subfigure}
\caption{Neutrino slice status for all fully contained tracks in $\nu_{\mu}$ CC events. Error bars shown on MicroBooNE data represent only statistical uncertainty.}
\label{fig:slice_presence_preselection}
\end{figure*}

\begin{enumerate}
    \item Signal Proton: Track belongs to a secondary proton produced by a final state neutron through neutron-inelastic scatter (final states (3), (4), and (6) in Tab.~\ref{table:NeutronFinalState}).
    \item Neutron Inelastic (NI) Proton (non-signal): Track belongs to a tertiary proton that has been produced by a neutrino-generated, non-final state neutron through neutron-inelastic scatter (final states (3), (4), and (6) in Tab.~\ref{table:NeutronFinalState}).
    \item Primary Particle: Track belongs to a final state particle produced directly in the neutrino interaction.
    \item Proton Inelastic (PI) Proton: Track belongs to a secondary proton produced through proton-inelastic scatter.
    \item Other Particle: Track belongs to any other simulated particle. 
    \item Dirt: Track originates from a neutrino interaction that occurred in the detector hall or dirt surrounding the detector complex.
    \item Cosmics and MC EM: Track belongs to either simulated electromagnetic (EM) activity or data cosmic activity overlaid on top of simulated neutrino event.
    \item BNB Beam-Off Data: Cosmic-induced backgrounds, measured with data taken when the beam was off.
    \item BNB Beam-On Data.
\end{enumerate}

\begin{figure*}[htb!]
\centering
\begin{subfigure}{.98\textwidth}
    \centering
    \includegraphics[width=.9\linewidth]{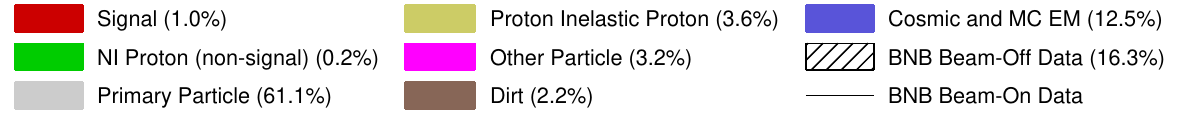}
\end{subfigure}
\begin{subfigure}{0.49\textwidth}
    \includegraphics[width=\linewidth]{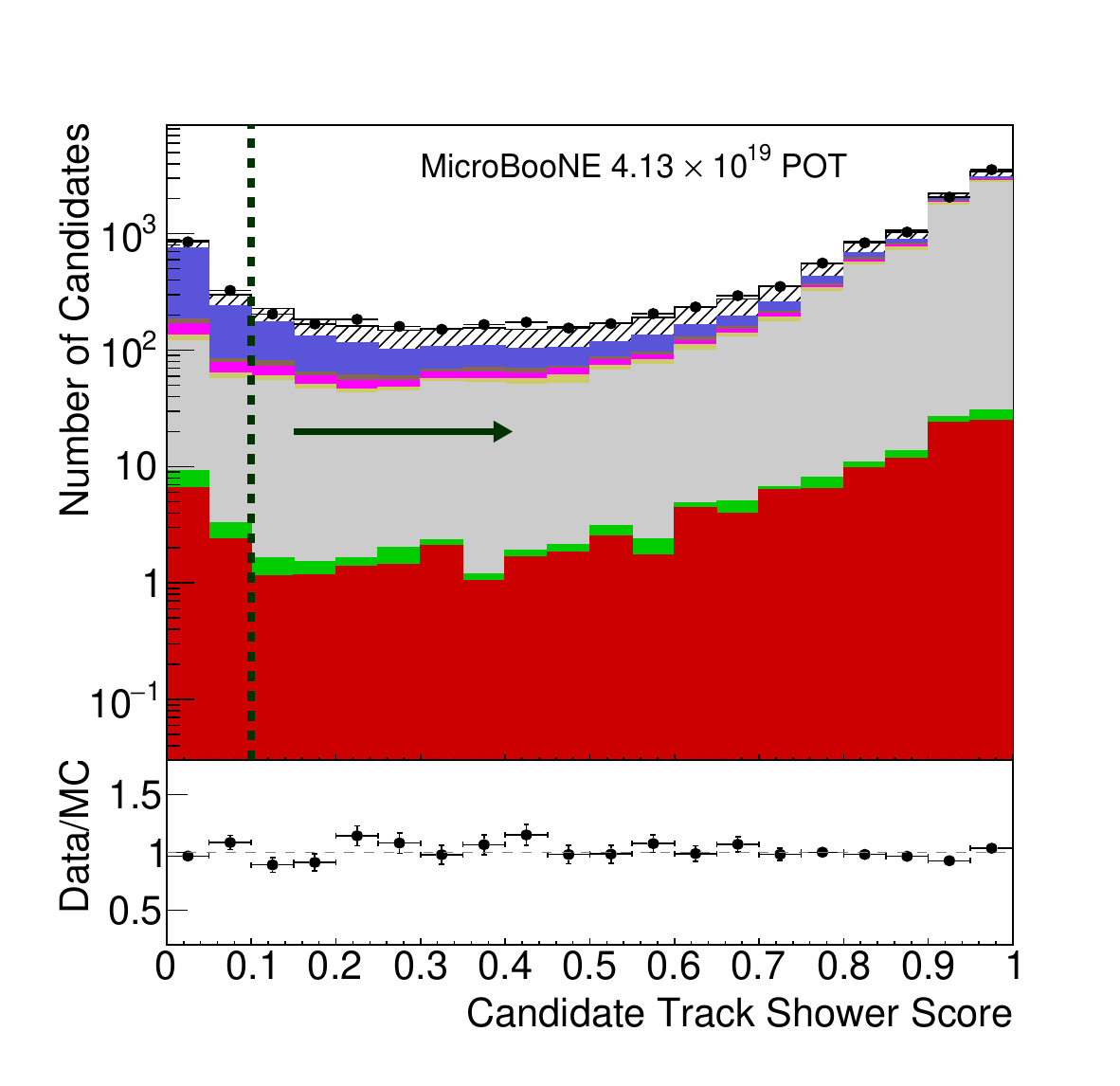}
    \caption{}
    \label{fig:run3_pfp_track_score_in_slice}
\end{subfigure}
\begin{subfigure}{0.49\textwidth}
    \includegraphics[width=\linewidth]{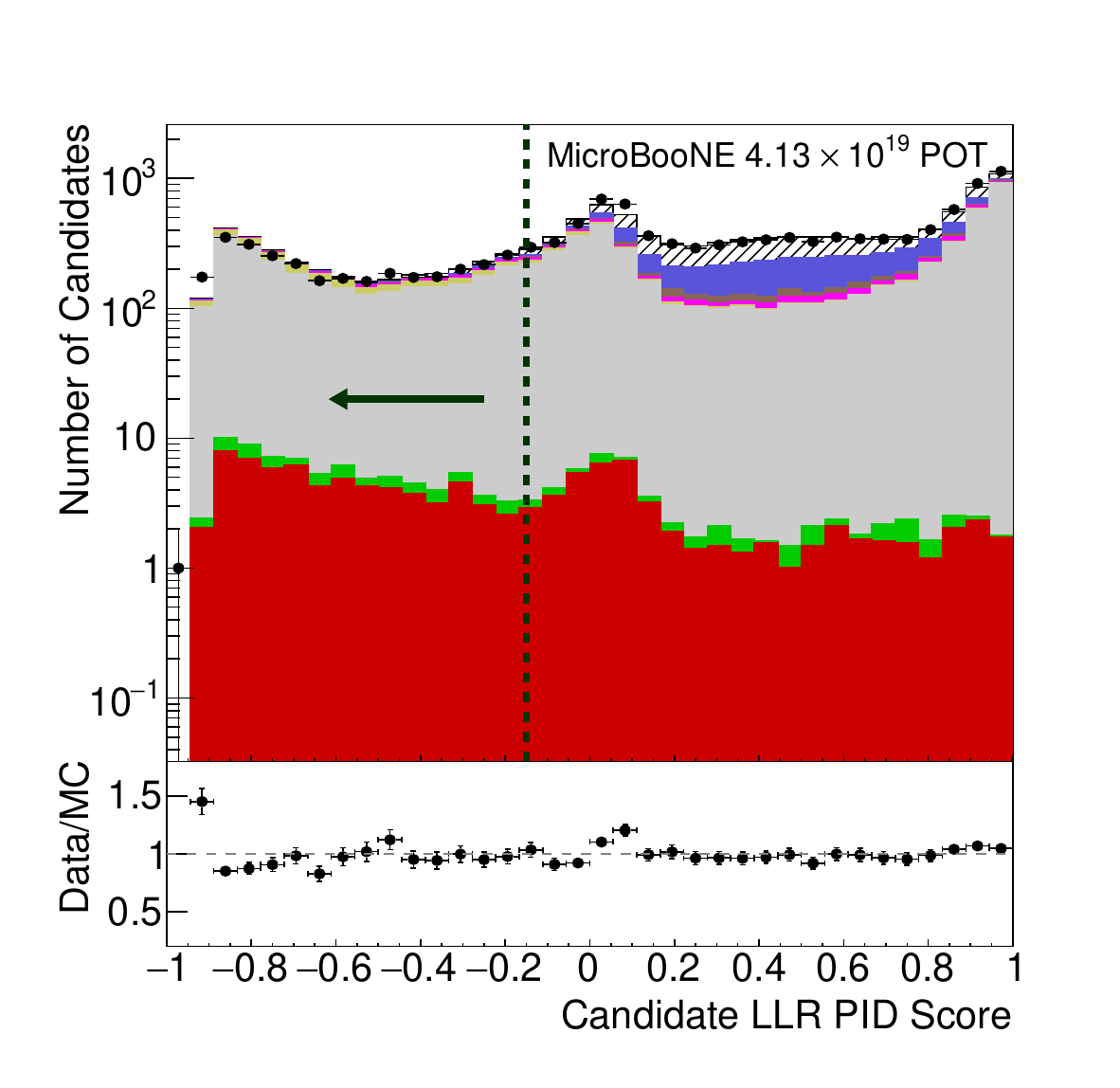}
    \caption{}
    \label{fig:run3_pfp_pidscore_in_slice}
\end{subfigure}
\begin{subfigure}{0.49\textwidth}
    \includegraphics[width=\linewidth]{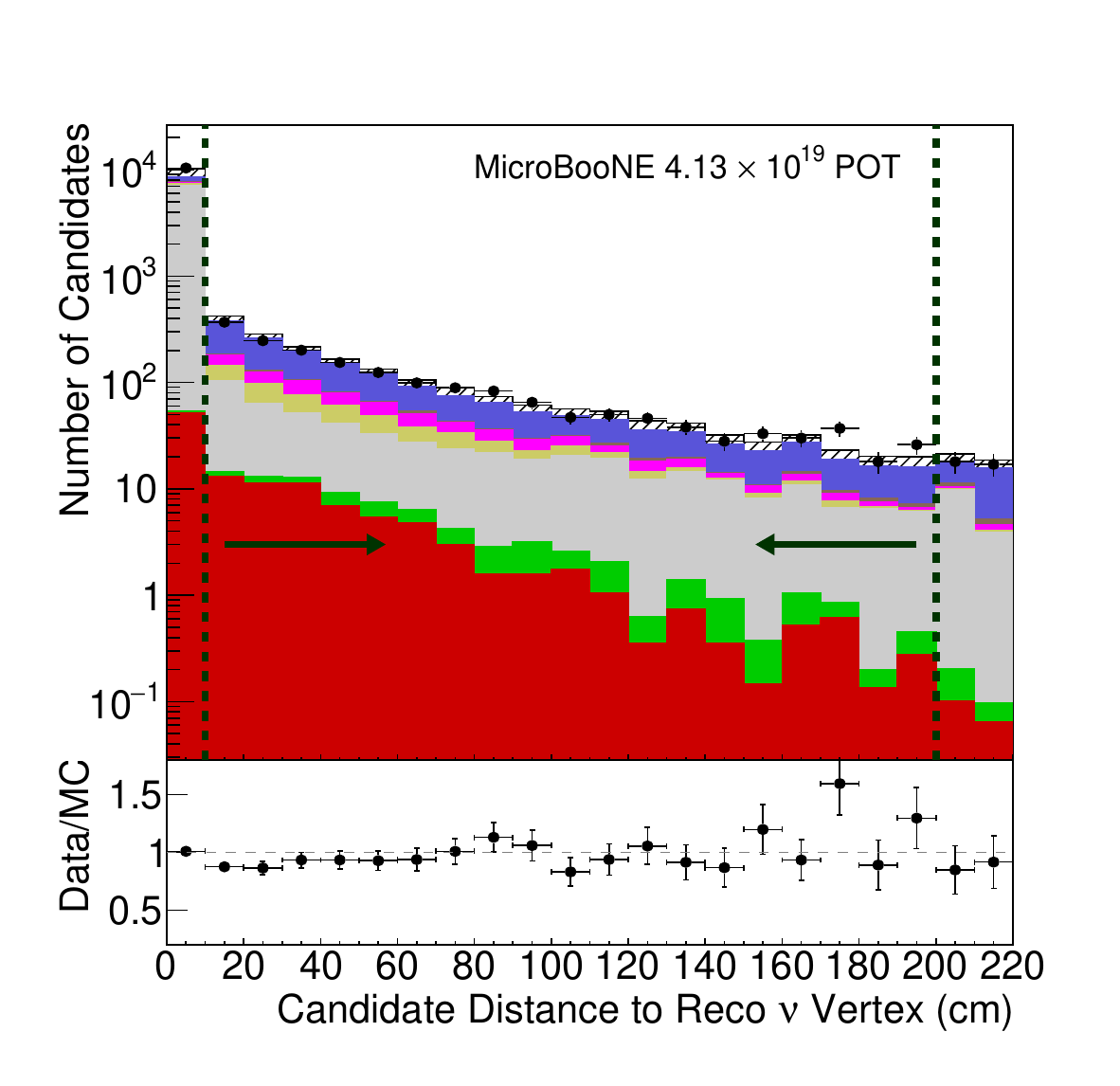}
    \caption{}
    \label{fig:run3_pfp_vtx_disp_in_slice}
\end{subfigure}
\begin{subfigure}{0.49\textwidth}
    \includegraphics[width=\linewidth]{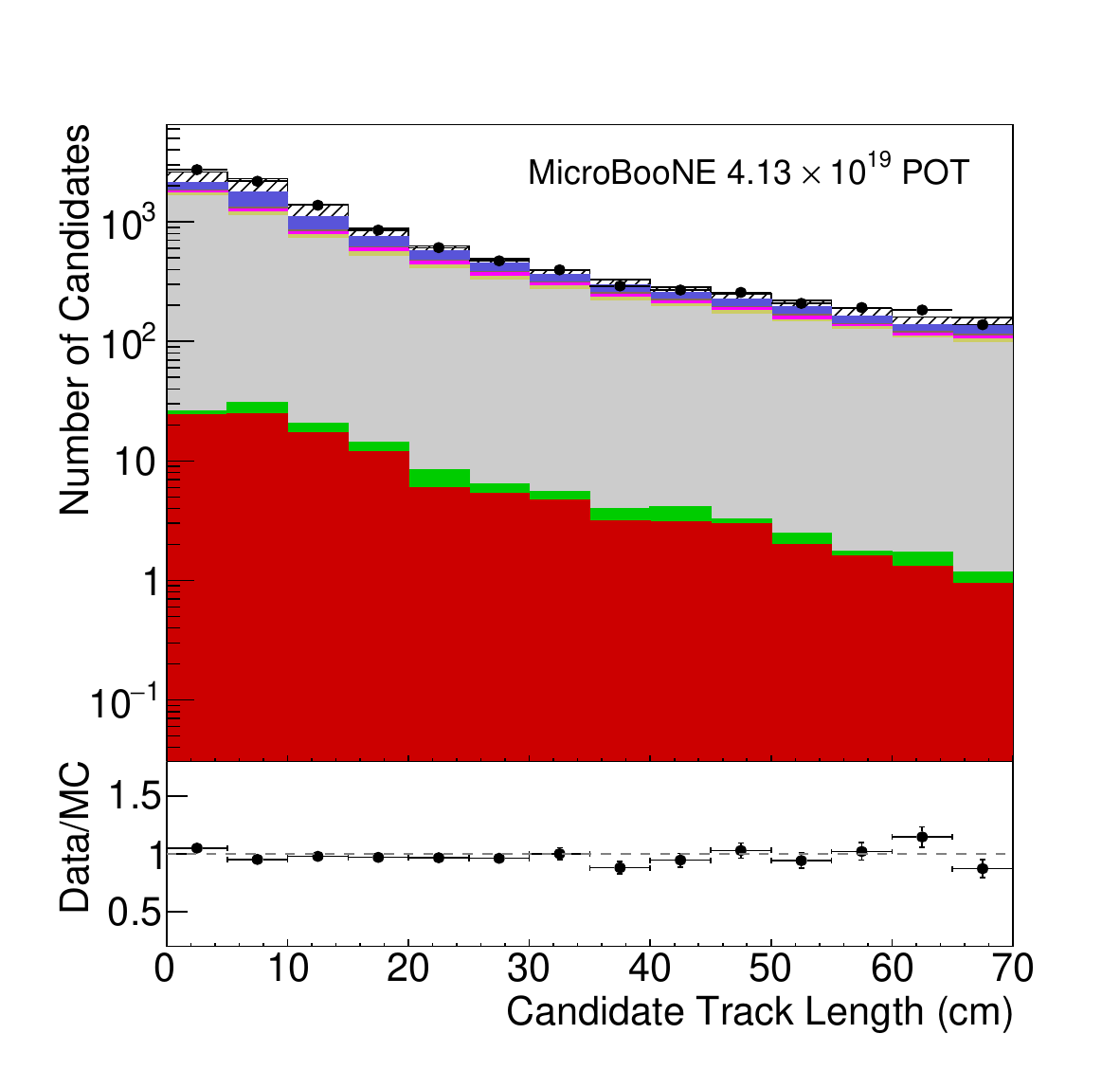}
    \caption{}
    \label{fig:run3_pfp_len_in_slice}
\end{subfigure}
\caption{All fully contained tracks in $\nu_{\mu}$ CC events that are reconstructed in the neutrino slice. (a) track/shower SVM score, (b) LLR PID score, (c) displacement between the candidate and the reconstructed neutrino vertex, (d) candidate track length.}
\label{fig:in_slice_plots}
\end{figure*}

All figures showing MicroBooNE simulation have been normalized to the data sample exposure used in this study ($4.13 \times 10^{19}$\,POT). BNB Beam-Off Data has been normalized to the total number of triggers in the BNB Beam-On Data sample. We define the integrated efficiency of the selection as the ratio between the number of true neutrons tagged from daughter proton tracks to the total number of true final state neutrons in the MC sample.
The selection purity is defined as the ratio of the number of signal proton tracks to the number of selected candidate tracks. 
NI non-signal protons are a background of particular interest.
Neutrino-generated neutrons that are created from the inelastic, secondary scatters of final state hadrons (primarily protons and pions) with argon nuclei can then scatter again and create tertiary protons.
This background is effectively indistinguishable from our signal, however, identifying these neutrons is useful.
The presence of any neutrino-generated neutrons indicates that there is missing energy, biasing the neutrino energy reconstruction.
Therefore, we will also report the NI produced proton purity, defined here as the ratio of the number of signal and NI proton tracks to the number of selected candidate tracks.

We require that neutron induced proton track candidates be reconstructed as part of the neutrino interaction by Pandora.
While this requirement rejects the majority ($62.0\%$) of our signal, it helps reject $95.6\%$ of our background. This information is shown in Fig.~\ref{fig:slice_presence_preselection}, where the large background rejection achieved by discarding tracks not reconstructed as part of the interaction (“Not in $\nu$ slice”) can be seen.
Future iterations of the analysis will attempt to recover these protons.

\begin{figure*}[htb!]
\centering
\begin{subfigure}{.98\textwidth}
    \centering
    \includegraphics[width=.9\linewidth]{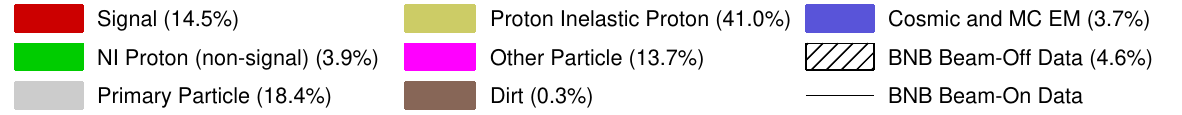}
\end{subfigure}
\begin{subfigure}{0.49\textwidth}
    \includegraphics[width=\linewidth]{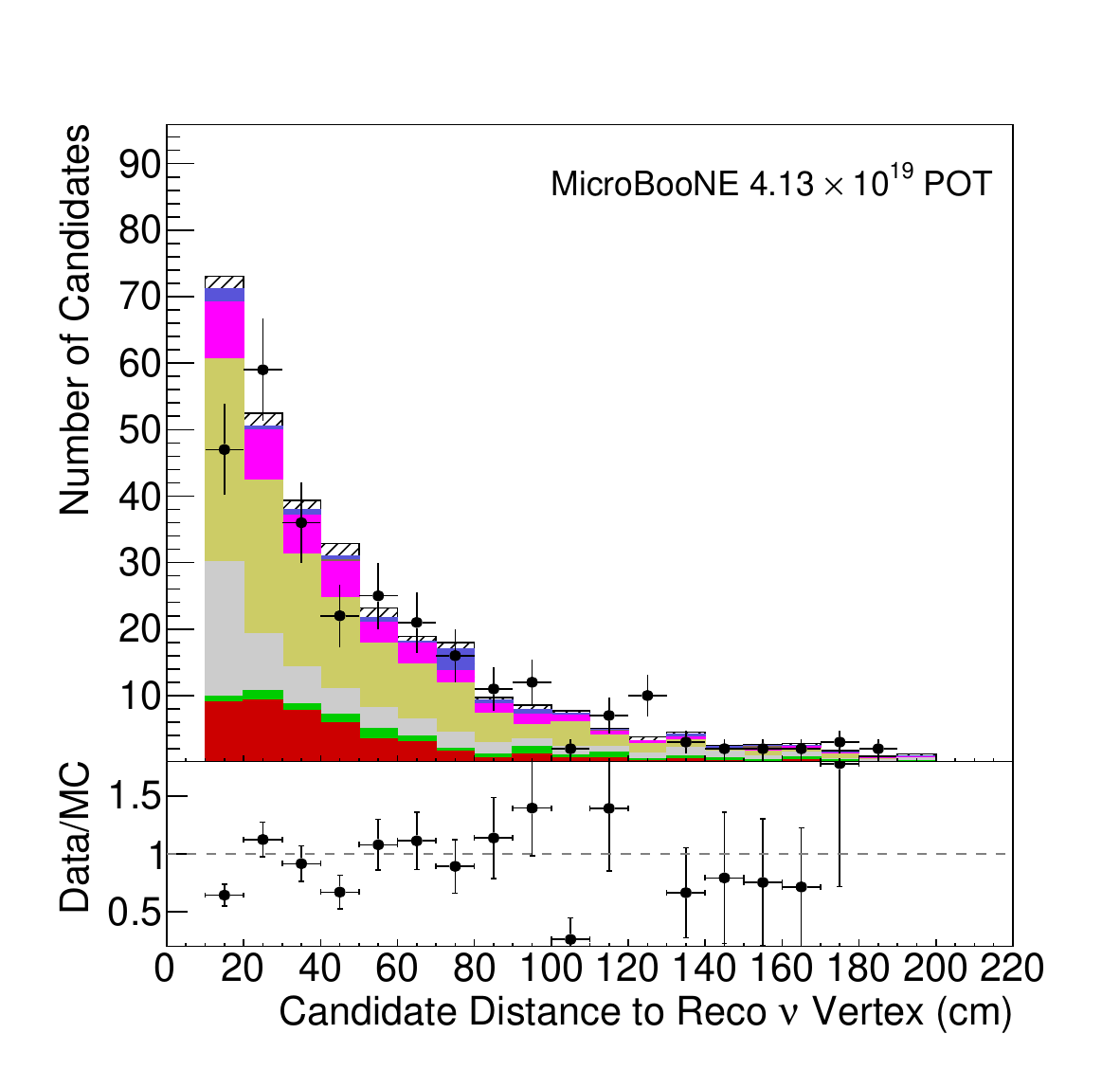}
    \caption{}
    \label{fig:PID_base_cut_vtx}
\end{subfigure}
\begin{subfigure}{0.49\textwidth}
    \includegraphics[width=\linewidth]{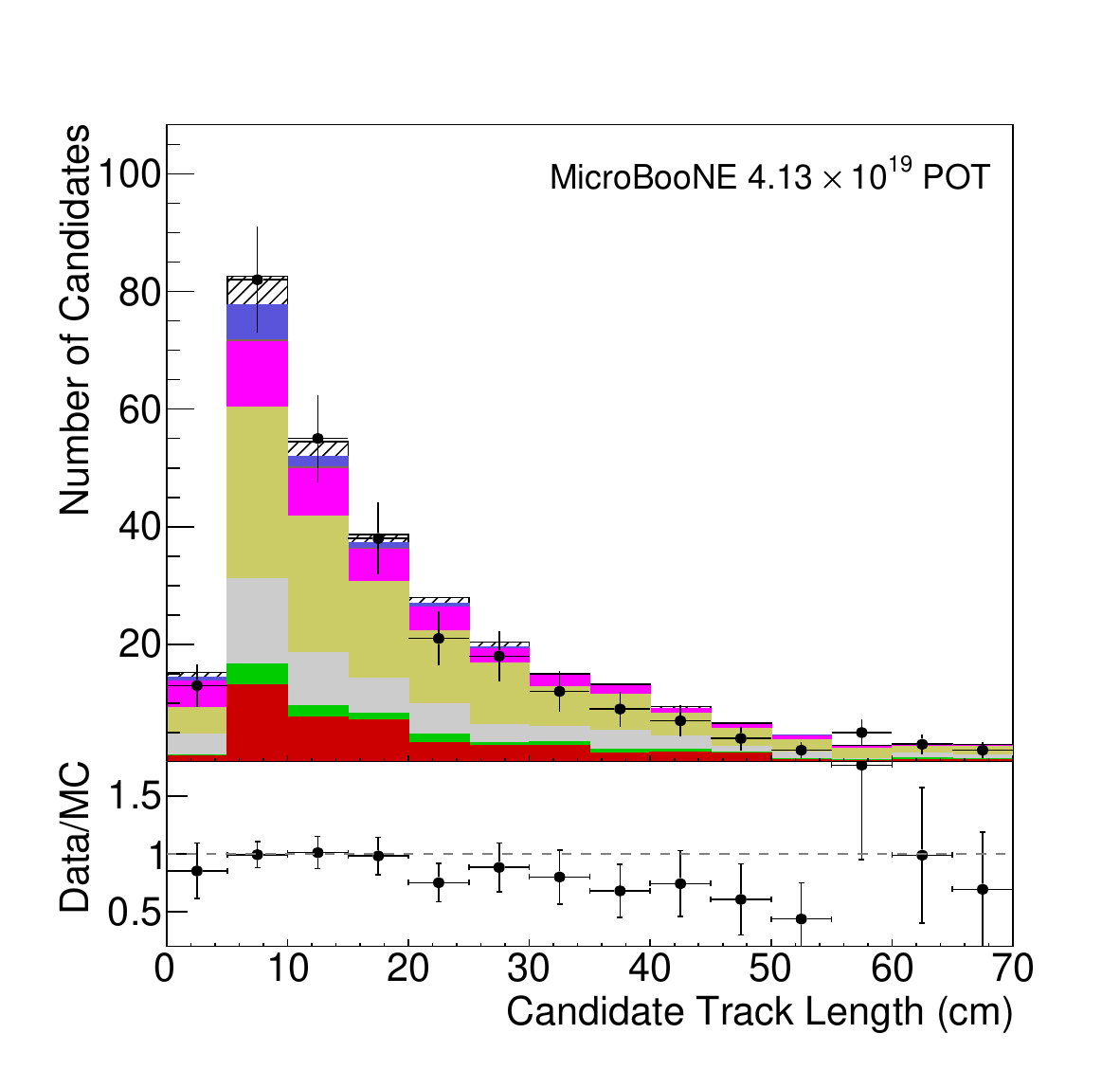}
    \caption{}
    \label{fig:PID_base_cut_len}
\end{subfigure}
\begin{subfigure}{0.49\textwidth}
    \includegraphics[width=\linewidth]{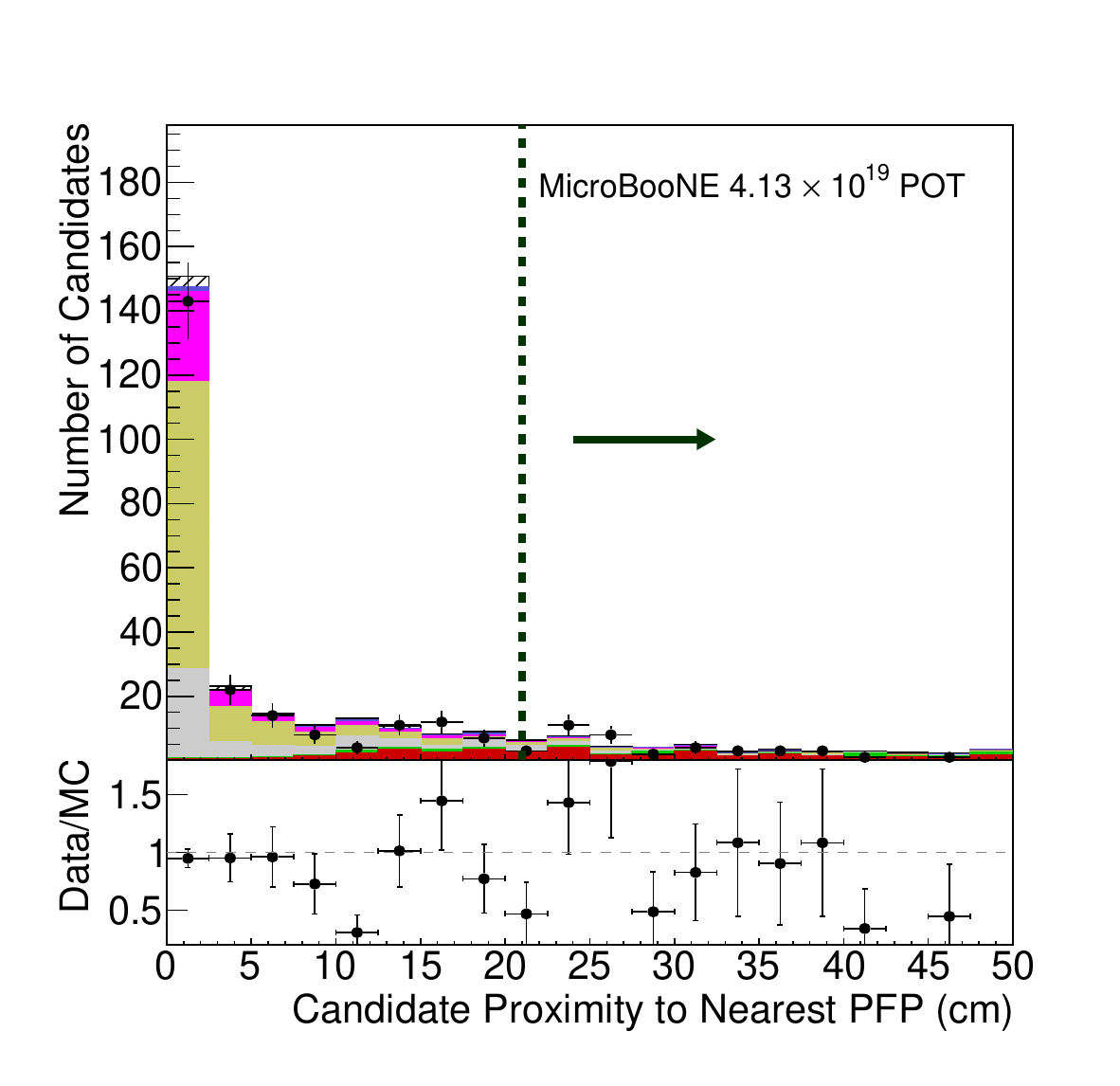}
    \caption{}
    \label{fig:run3_pfp_parent_prox_pid_trkshwr_disp_cuts}
\end{subfigure}
\begin{subfigure}{0.49\textwidth}
    \includegraphics[width=\linewidth]{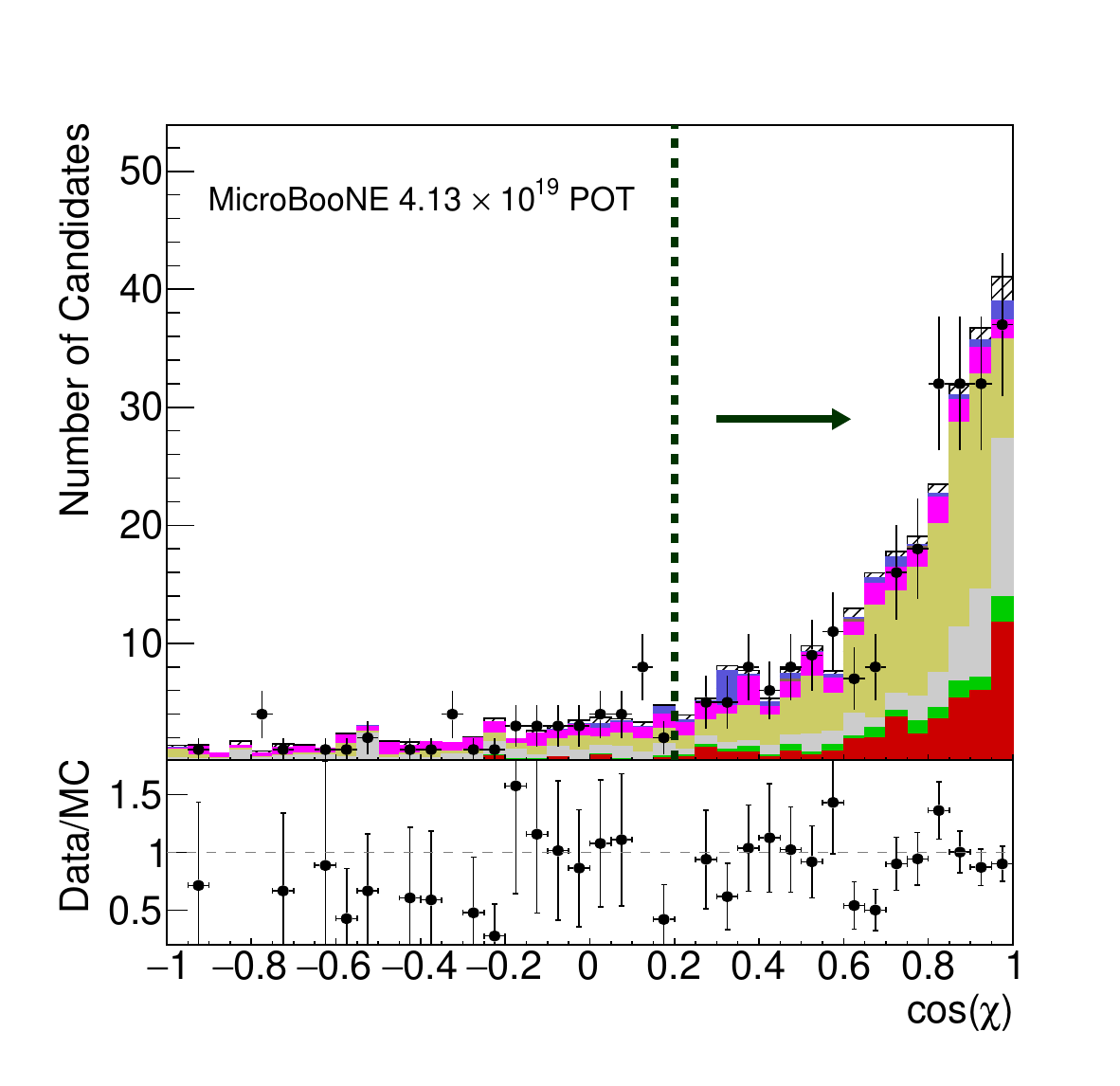}
    \caption{}
    \label{fig:run3_pfp_direction_pid_trkshwr_disp_cuts}
\end{subfigure}
\caption{Fully contained, neutrino-slice tracks in $\nu_{\mu}$ CC events with LLR PID Score $<$ $-0.15$, track/shower score $> 0.1$, and that lie between $10$ and $200$\,cm of the reconstructed neutrino vertex. (a) Candidate displacement from the reconstructed neutrino vertex, (b) candidate track length, (c) Candidate proximity to nearest PFP, (d) and PFP relative direction or cos($\chi$).}
\end{figure*}

Once PFPs that are not included in the neutrino slice are eliminated, the leading backgrounds are primary, BNB Beam-Off, Cosmic/EM MC tracks, and PI protons (Fig.~\ref{fig:in_slice_plots}).
The primary and PI proton backgrounds are both susceptible to uncertainties on GEANT4 modeling of particle propagation and reinteraction in the LAr.
The MC EM background is sensitive to uncertainties from our particle identification which relies on d$E$/d$x$ measurements. The dominant uncertainty on this is from electron-ion recombination \cite{recombination}.

At this point, both the Cosmic and BNB Beam-Off backgrounds are predominantly either muons or EM showers from muons. In order to eliminate them, we leverage two reconstruction variables.
The first variable is the track/shower score, which is the output of an SVM in Pandora that determines if the PFP appears more track-like or shower-like (Fig.~\ref{fig:run3_pfp_track_score_in_slice}).
Tracks have scores closer to 1, while showers have scores closer to 0.
The second reconstruction variable is the log-likelihood ratio particle identification (LLR PID) score (Fig.~\ref{fig:run3_pfp_pidscore_in_slice}) and is calculated using MicroBooNE-specific software that runs after Pandora completes its general reconstruction \cite{LLRPID_Paper}.
The score identifies tracks as more muon-like (closer to 1) or more proton-like (closer to -1). 
Both scores have been extensively vetted using MicroBooNE data and have been shown to be minimally impacted by systematic uncertainties \cite{GardinerNote, TKI_PRD}.

We remove most primary tracks by measuring the displacement between the candidate start point and the reconstructed neutrino vertex (Fig.~\ref{fig:run3_pfp_vtx_disp_in_slice}).
After correcting for electric field distortions in our calibrations \cite{SCE_microboone}, both this variable and the proton candidate track length (Fig.~\ref{fig:run3_pfp_len_in_slice}) are minimally impacted by the space charge effect, with the largest effect when either the neutrino vertex or neutron candidate are close to the edge of the TPC.

The cuts on track/shower score, LLR PID score, and reconstructed vertex separation were optimized to maximize the product of the integrated neutron detection efficiency and signal purity.
The selection criteria are:
\begin{itemize}
    \item track/shower score $> 0.1$ (Fig.~\ref{fig:run3_pfp_track_score_in_slice}),
    \item LLR PID score $< -0.15$ (Fig.~\ref{fig:run3_pfp_pidscore_in_slice}),
    \item vertex separation $10<d<200$\,cm (Fig.~\ref{fig:run3_pfp_vtx_disp_in_slice}).
\end{itemize}
The vertex separation and track length are shown in Fig.~\ref{fig:PID_base_cut_vtx} and Fig.~\ref{fig:PID_base_cut_len} for the surviving tracks.
At this point, the largest backgrounds are proton-inelastic and primary protons.
Unlike their neutron counterpart, the initiating proton in proton-inelastic tracks is visible in the detector.
The secondary proton in proton inelastic interactions will inherently be in close proximity to the parent proton, thereby differentiating it from our signal.
Assuming no nearby cosmic or primary PFPs, neutron produced proton tracks are isolated in space.
We measure the shortest distance between either end of the secondary proton candidate, and the start and end points of every other PFP in the detector.
The minimum of the 4N calculations---with N+1 being the number of PFPs in the detector including the current proton candidate---is the nearest PFP proximity.
The variable ``candidate proximity to the nearest PFP" gives good discriminating power between signal and these two proton backgrounds and is shown in Fig.~\ref{fig:run3_pfp_parent_prox_pid_trkshwr_disp_cuts}.

Assuming the neutron direction is given by a straight line from the reconstructed neutrino vertex and the start of the proton candidate, we measure the cosine of the relative angle ($\cos (\chi)$) between the neutron direction and the candidate proton direction as seen in Fig.~\ref{fig:run3_pfp_direction_pid_trkshwr_disp_cuts}.
The candidate direction is defined as the vector between the reconstructed start and end positions of the candidate track.
At this point in the event selection, the remaining cosmic, BNB Beam-Off, and Dirt tracks are likely neutron-induced protons.
Neutrino-generated tracks should be forward-peaked in this variable due to momentum conservation whereas the other backgrounds should not.
As the relative direction is specifically designed to reduce the neutron background that enters from outside the detector, it is particularly susceptible to the uncertainty on modeling neutrino interactions in the dirt around the detector \cite{PhysRevD.105.112005}.
However, the overall effect of dirt uncertainties should be small as the dirt background is very small.
The selection criteria for both of the physics variables are:
\begin{itemize}
    \item candidate PFP proximity $> 21$\,cm,
    \item candidate track $\cos\chi > 0.2$.
\end{itemize}
Both the PFP proximity and relative direction will also be minimally affected by the space charge effect in LArTPCs.

\begin{figure*}[htb!]
    \centering
    \begin{subfigure}{.98\textwidth}
        \centering
        \includegraphics[width=.9\linewidth]{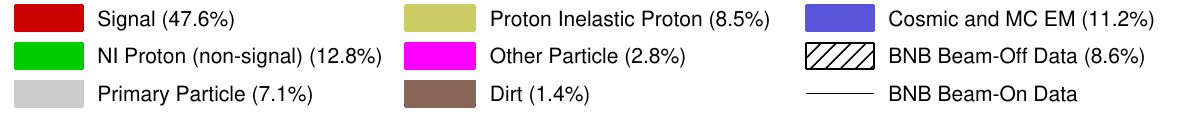}
    \end{subfigure}
    \begin{subfigure}[t]{0.49\textwidth}
        \centering
        \includegraphics[width=1.\linewidth]{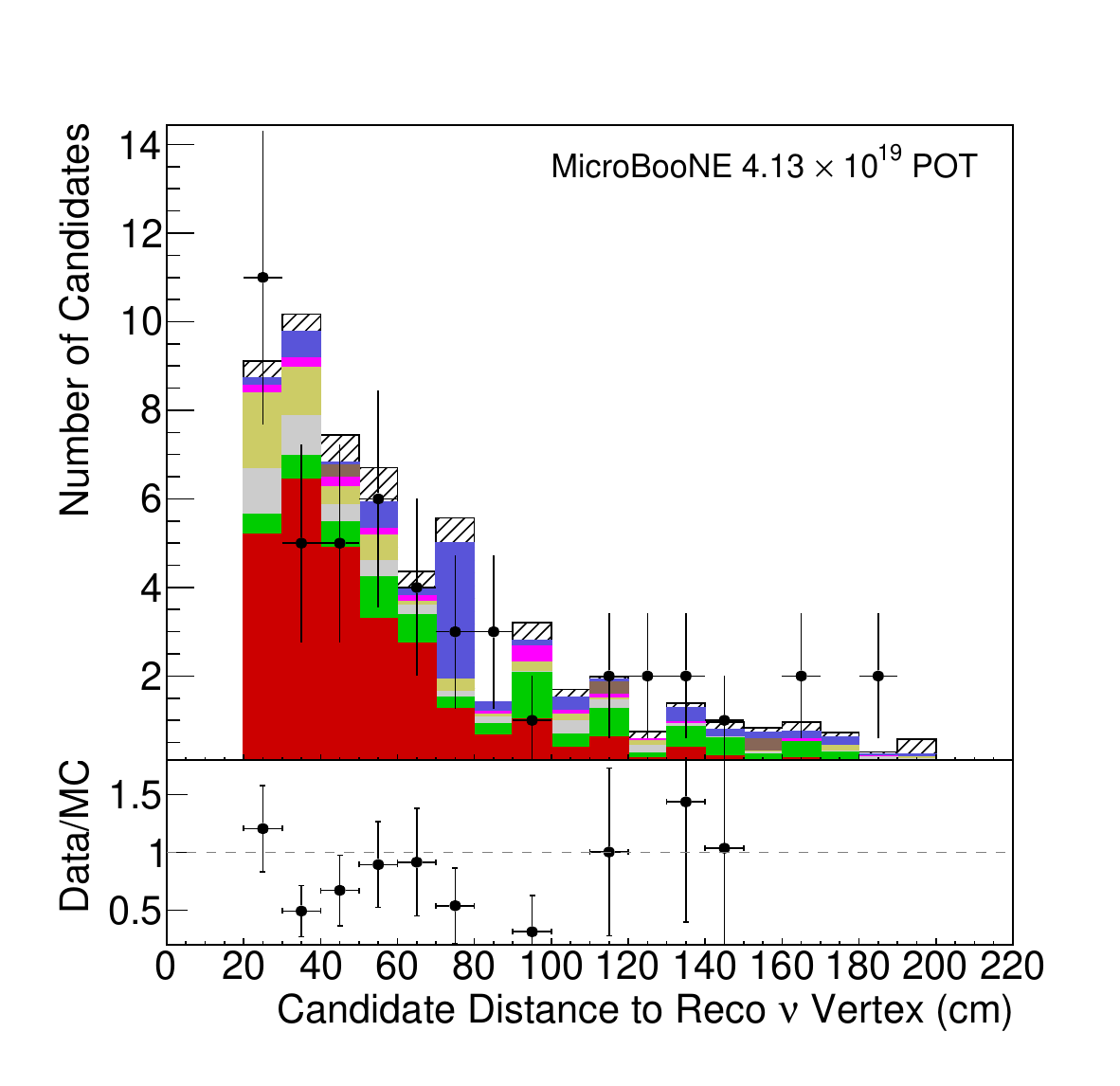}
        \caption{}
        \label{fig:final_vtx}
    \end{subfigure}%
    ~ 
    \begin{subfigure}[t]{0.49\textwidth}
        \centering
        \includegraphics[width=1.\linewidth]{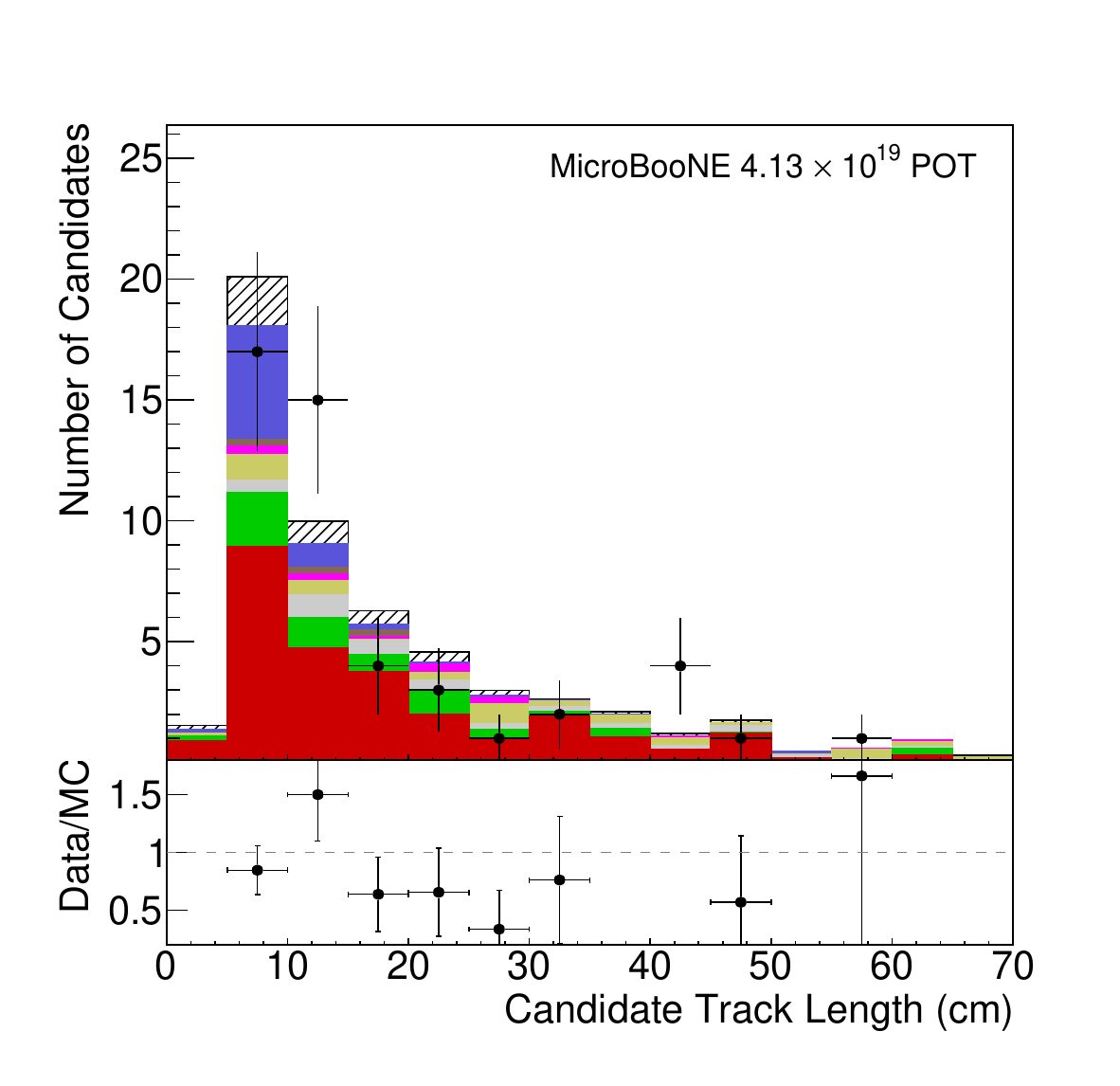}
        \caption{}
        \label{fig:final_len}
    \end{subfigure}
    \caption{(a) Final candidate displacement from the reconstructed neutrino vertex. (b) Final candidate track length.}
\end{figure*}

\begin{figure*}[htb!]
    \centering
    \begin{subfigure}{.98\textwidth}
        \includegraphics[width=.9\linewidth]{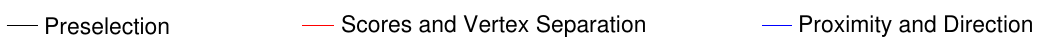}
    \end{subfigure}
    \begin{subfigure}[t]{0.49\textwidth}
        \centering
        \includegraphics[width=1.\linewidth]{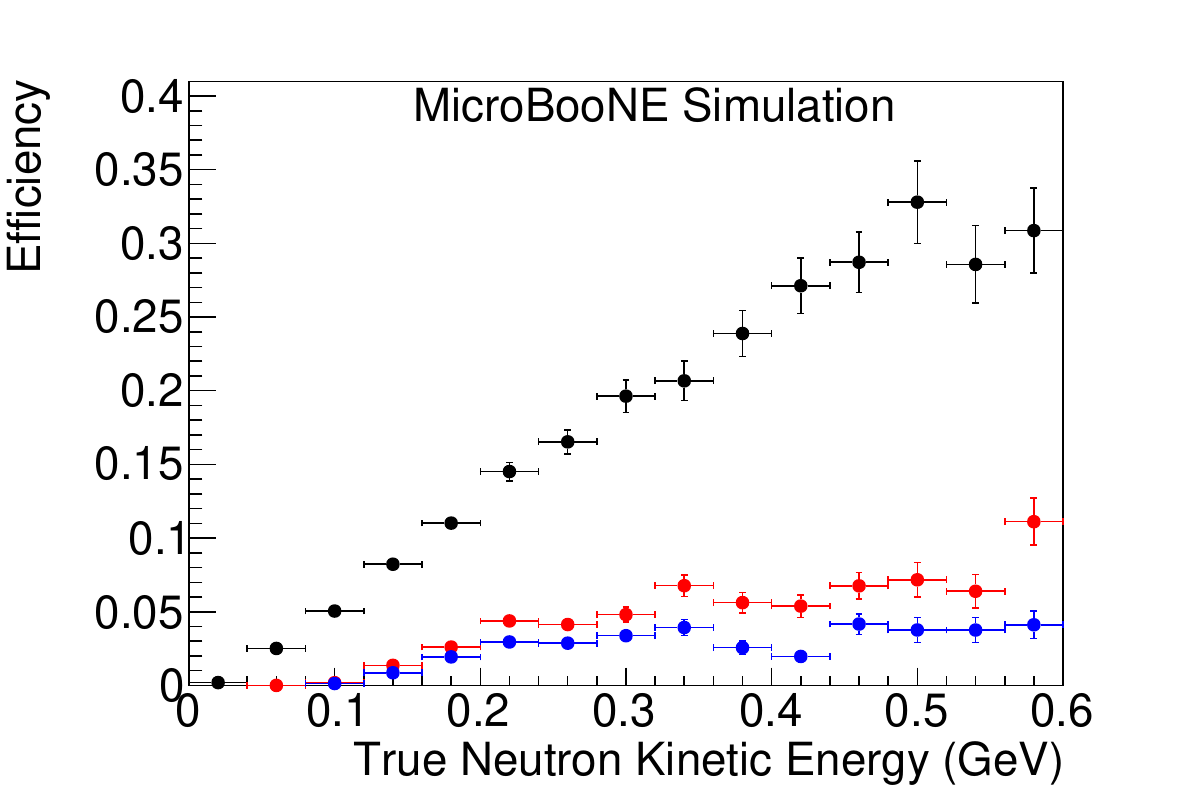}
        \caption{}
        \label{fig:neutron_efficiency_KE}
    \end{subfigure}
    \begin{subfigure}[t]{0.49\textwidth}
        \centering
        \includegraphics[width=1.\linewidth]{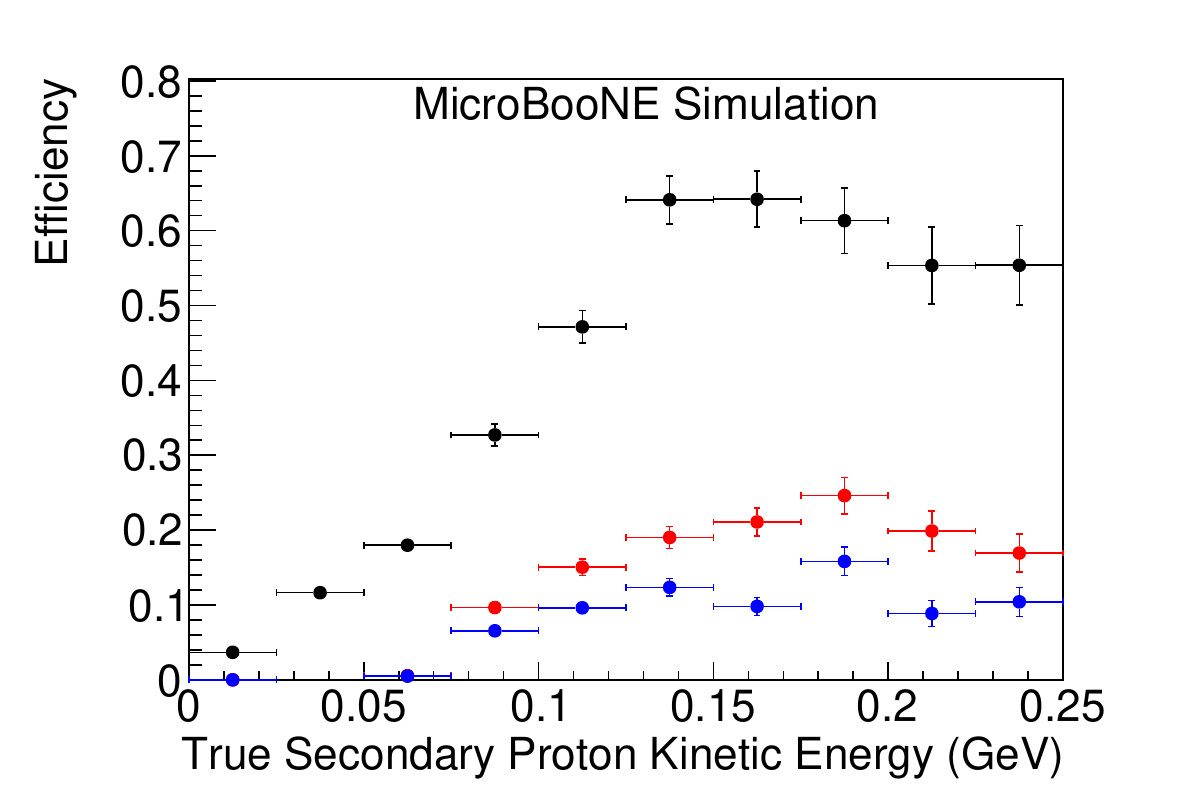}
        \caption{}
        \label{fig:secondary_proton_efficiency_KE}
    \end{subfigure}
    \begin{subfigure}[t]{0.49\textwidth}
        \centering
        \includegraphics[width=1.\linewidth]{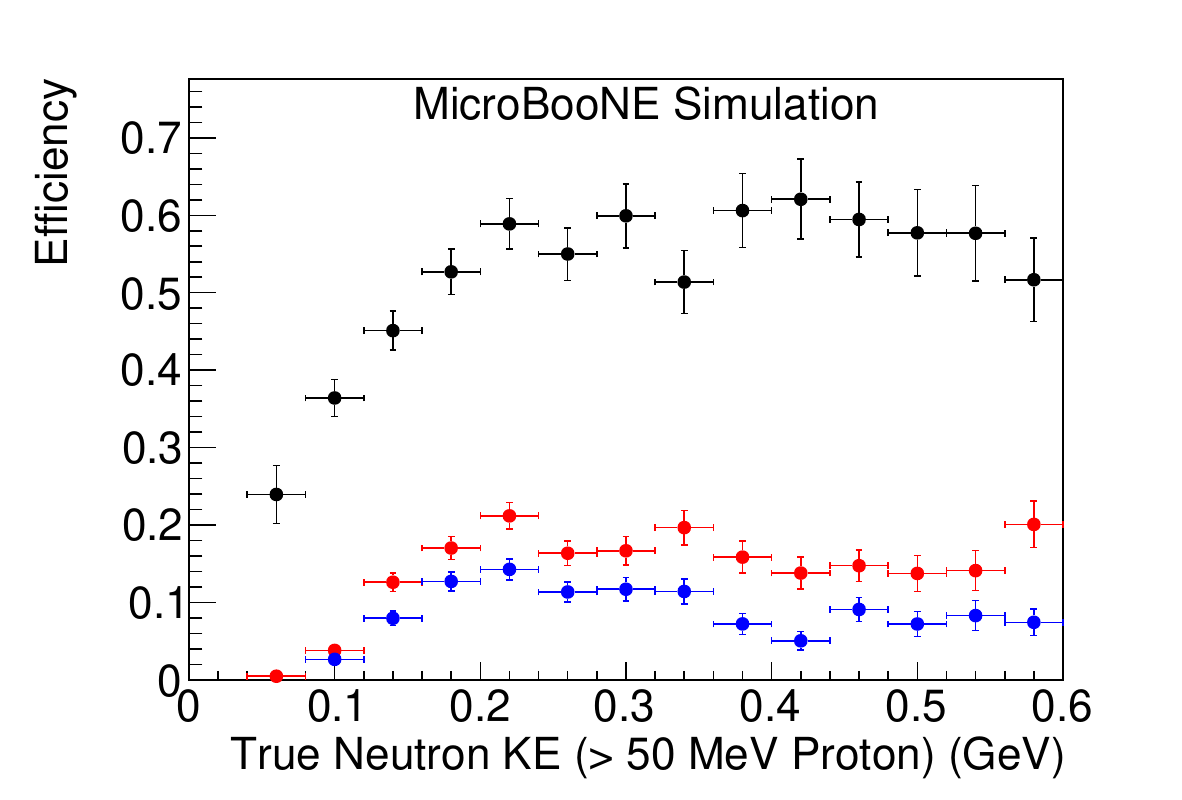}
        \caption{}
        \label{fig:neutron_efficiency_KE_50MeV_p}
    \end{subfigure}
    \begin{subfigure}[t]{0.49\textwidth}
        \centering
        \includegraphics[width=1.\linewidth]{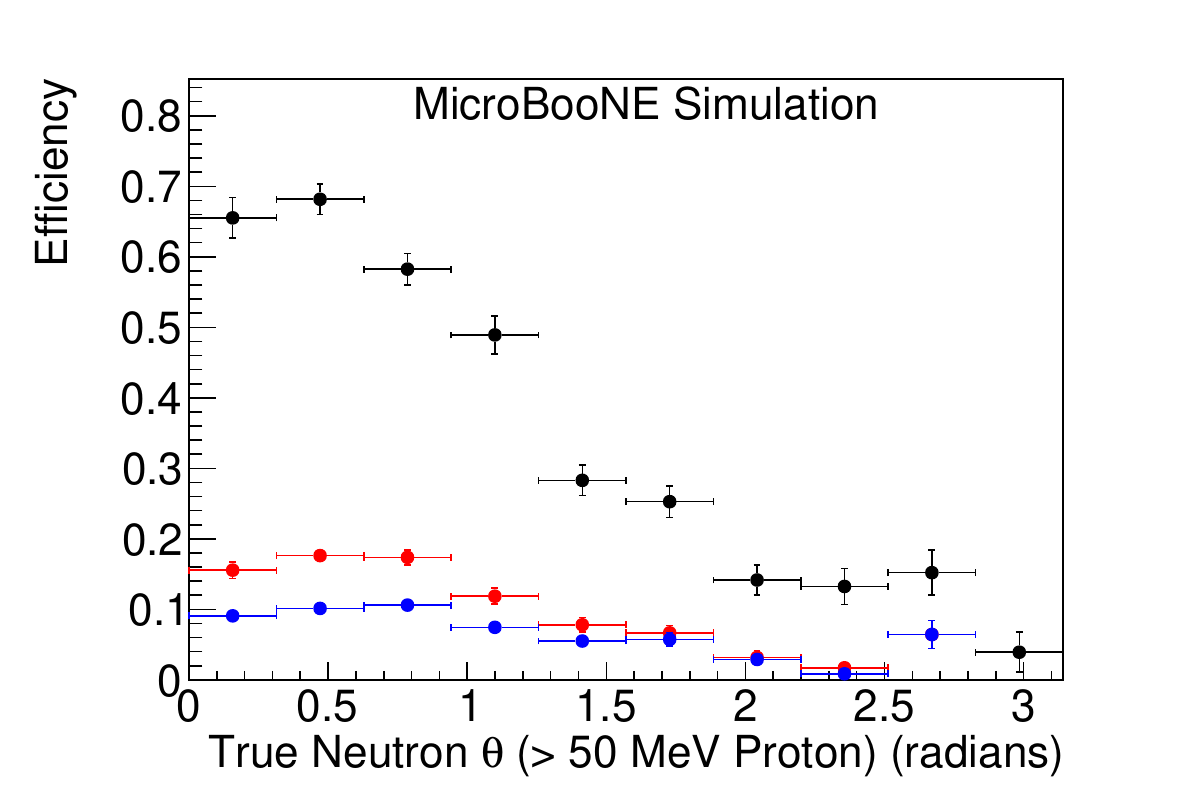}
        \caption{}
        \label{fig:neutron_efficiency_phi}
    \end{subfigure}
    \caption{(a) Neutron detection efficiency as a function of true neutron kinetic energy. (b) Secondary proton detection efficiency. (c) Neutron detection efficiency as a function of true neutron energy for neutrons that produce protons with kinetic energy 50 MeV or more. (d) Neutron detection efficiency as a function of neutron polar scattering angle for neutrons that produce protons with kinetic energy 50 MeV or more.}
    \label{fig:efficiencies}
\end{figure*}

\section{Event Selection Performance}
The reconstructed neutrino vertex displacement and track length for the final set of neutron-generated, secondary proton track candidates is shown in Fig.~\ref{fig:final_vtx} and~\ref{fig:final_len}.
These distributions demonstrate that the modeling of neutron inelastic scatters in MicroBooNE agrees with data within the sizeable statistical uncertainties in this sample.

The final sample of neutron candidates is 48\% pure primary neutrons, and 60\% pure neutrino-induced neutrons (including non-signal neutrons), with the largest non-neutron backgrounds being cosmic particles, proton-inelastic protons, and primary particles (also primarily protons).
Due to our use of data overlays, it is not possible to know exactly what type of cosmic particles are being selected, but some fraction will be cosmogenic neutrons scattering close to the neutrino interaction.

Figure~\ref{fig:neutron_efficiency_KE} shows the neutron detection efficiency as a function of true neutron kinetic energy at different stages in the event selection (error bars in Fig.~\ref{fig:efficiencies} represent only statistical uncertainty).
The final efficiency is negligible below neutron energies of around 100\,MeV, rising to an approximately flat efficiency of around 3\% above 250\,MeV.
The secondary proton detection efficiency as a function of true proton kinetic energy (Fig.~\ref{fig:secondary_proton_efficiency_KE}) shows that a significant source of inefficiency is failing to reconstruct and select protons below 50\,MeV kinetic energy.
This efficiency measures the detection rate of neutron-inelastically produced protons that start in the TPC.
There are a small number of neutrons that create multiple secondary proton tracks that pass all event selection cuts.
These neutrons are counted once when determining the neutron efficiency, however Fig.~\ref{fig:secondary_proton_efficiency_KE} counts each secondary proton independently.
Due to this proton threshold, we report an additional neutron detection efficiency as a function of neutron kinetic energy for neutrons that create a proton in the TPC with more than 50\,MeV of true kinetic energy in Fig.~\ref{fig:neutron_efficiency_KE_50MeV_p}.
Lastly, we report the neutron detection efficiency as a function of neutron polar scattering angle (defined as the angle from the beam direction) which shows better detection efficiency for neutrons that are more forward going (Fig.~\ref{fig:neutron_efficiency_phi}).
This higher detection efficiency at low polar angle is due to several reasons: forward-going neutrons tend to have higher energies leading to higher efficiency; the detector geometry means forward-going neutrons have a lower chance of exiting the detector before scattering; and the slicing performed by Pandora has some dependence on direction, therefore forward-going tracks are more likely to be classified as neutrino-like.

While the efficiency integrated across all neutron energies is low, the majority of neutrons produced are very low energy.
For neutrons that create protons in the TPC with 50\,MeV of kinetic energy or more, the integrated efficiency is $8.4\%$.

The secondary proton energy threshold of approximately 50\,MeV is consistent with the threshold for primary protons in charged-current interactions \cite{PhysRevD.102.112013}.
However, due to the lack of hits from other particles in the same location, these isolated charge depositions can be reconstructed at significantly lower energies in principle.  
Further work to lower the proton threshold, including leveraging newly developed low-threshold reconstruction capabilities \cite{RadonPaper, BiPoPaper}, would lead to significant efficiency gains.
Additionally, the selection could be expanded to consider protons not reconstructed as part of the neutrino slice -- around 60\% of neutron-induced protons in our sample.

\section{Summary and Conclusions}
\label{sec:summary}

We have demonstrated the ability to identify neutrons produced in neutrino interactions through the production of secondary protons from neutron interactions separated from the neutrino vertex.
The sample of neutron candidates identified is 60\% pure, with 48\% purity when only considering primary neutrons.
While the integrated efficiency remains low, there remain prospects for improvement.
Additionally, the efficiency is lowest for low-energy neutrons which contribute less to neutrino energy biases, and highest for neutrons above 250\,MeV kinetic energy, which could represent significant biases to neutrino energy estimates.

The methods outlined in this paper would be applicable to any LArTPC and allow for the measurement of neutrino-induced neutron production in MicroBooNE, the SBN experiments, or DUNE in the future.
Additionally, this method can be used to provide statistical separation of neutrino and antineutrino events in a non-magnetized LArTPC, and statistical separation of a sample of events known to have biased neutrino energy estimators which may reduce uncertainties in an oscillation measurement.

\begin{acknowledgements}
%This document was prepared by the MicroBooNE collaboration using the resources of the Fermi National Accelerator Laboratory (Fermilab), a U.S.\ Department of Energy, Office of Science, HEP User Facility. Fermilab is managed by Fermi Research Alliance, LLC (FRA), acting under Contract No.\ DE-AC02-07CH11359.  MicroBooNE is supported by the following: the U.S.\ Department of Energy, Office of Science, Offices of High Energy Physics and Nuclear Physics; the U.S. National Science Foundation; the Swiss National Science Foundation; the Science and Technology Facilities Council (STFC), part of the United Kingdom Research and Innovation; the Royal Society (United Kingdom); and The European Union’s Horizon 2020 Marie Sklodowska-Curie Actions. Additional support for the laser calibration system and cosmic ray tagger was provided by the Albert Einstein Center for Fundamental Physics, Bern, Switzerland. We also acknowledge the contributions of technical and scientific staff to the design, construction, and operation of the MicroBooNE detector as well as the contributions of past collaborators to the development of MicroBooNE analyses, without whom this work would not have been possible.
This document was prepared by the MicroBooNE collaboration using the
resources of the Fermi National Accelerator Laboratory (Fermilab), a
U.S. Department of Energy, Office of Science, HEP User Facility.
Fermilab is managed by Fermi Research Alliance, LLC (FRA), acting
under Contract No. DE-AC02-07CH11359.  MicroBooNE is supported by the
following: 
the U.S. Department of Energy, Office of Science, Offices of High Energy Physics and Nuclear Physics; 
the U.S. National Science Foundation; 
the Swiss National Science Foundation; 
the Science and Technology Facilities Council (STFC), part of the United Kingdom Research and Innovation; 
the Royal Society (United Kingdom); 
the UK Research and Innovation (UKRI) Future Leaders Fellowship; 
and the NSF AI Institute for Artificial Intelligence and Fundamental Interactions. 
Additional support for 
the laser calibration system and cosmic ray tagger was provided by the 
Albert Einstein Center for Fundamental Physics, Bern, Switzerland. We 
also acknowledge the contributions of technical and scientific staff 
to the design, construction, and operation of the MicroBooNE detector 
as well as the contributions of past collaborators to the development 
of MicroBooNE analyses, without whom this work would not have been 
possible. 
For the purpose of open access, the authors have applied 
a Creative Commons Attribution (CC BY) public copyright license to 
any Author Accepted Manuscript version arising from this submission.
\end{acknowledgements}

\end{document}